\documentclass[10pt]{article}
\usepackage{fullpage}
\usepackage{amsmath}
\usepackage{amssymb}
\usepackage{amsfonts}
\usepackage[dvips]{epsfig}
\usepackage{color}

\def\##1{{\bf #1}}
\def\=#1{\underline{\underline{#1}}}

\def\+#1{\underline{\bf #1}}
\def\*#1{\underline{\underline{\bf #1}}}

\def\r#1{(\ref{#1})}
\def\l#1{\label{#1}}
\def\c#1{\cite{#1}}

\def\le{\left(}
\def\ri{\right)}
\def\les{\left[}
\def\ris{\right]}
\def\lec{\left\{}
\def\ric{\right\}}

\def\.{\cdot}

\def\epso{\epsilon_{\scriptscriptstyle 0}}

\def\muo{\mu_{\scriptscriptstyle 0}}

\def\ko{k_{\scriptscriptstyle 0}}

\def\eps{\epsilon}

\begin{document}

\begin{center}

{\bf {\LARGE Voigt waves in homogenized particulate composites
\vspace{3mm} \\ based on   isotropic dielectric components}}

\vspace{10mm} \large

 Tom G. Mackay\footnote{E--mail: T.Mackay@ed.ac.uk}\\
{\em School of Mathematics and
   Maxwell Institute for Mathematical Sciences\\
University of Edinburgh, Edinburgh EH9 3JZ, UK}\\
and\\
 {\em NanoMM~---~Nanoengineered Metamaterials Group\\ Department of Engineering Science and Mechanics\\
Pennsylvania State University, University Park, PA 16802--6812,
USA}\\

\end{center}

\vspace{4mm}

\normalsize

\begin{abstract}

Homogenized composite materials (HCMs) can support a singular form
of optical propagation, known as Voigt wave propagation, while their
component materials do not. This phenomenon was investigated for
biaxial HCMs arising from nondissipative isotropic dielectric
component materials. The biaxiality of these HCMs stems from the
oriented spheroidal shapes of the particles which make up  the
component materials. An extended version of the Bruggeman
homogenization formalism was used to investigate the influence of
component particle orientation, shape and size, as well as volume
fraction of the component materials, upon Voigt wave propagation.
Our numerical studies revealed that
  the directions in which
 Voigt waves  propagate is highly sensitive to the orientation
 of the component particles and to the volume fraction of the
 component materials, but  less sensitive to the shape of the
 component particles and  less sensitive still to the size of the
 component particles. Furthermore, whether or not such an HCM
 supports
 Voigt wave propagation at all is critically dependent upon the size of the
 component particles and, in certain cases, upon the volume
 fraction of the component materials.

\end{abstract}

\vspace{5mm} \noindent  {\bf Keywords:} Voigt waves, Bruggeman
homogenization formalism, optical singularities

\section{Introduction}

A composite material, comprising  random  distributions of two (or
more) particulate component materials, may be regarded as
effectively homogeneous provided that the particles which constitute
the component materials are much smaller than the wavelengths
involved \c{L96,Milton}. Depending upon the shape, size and
distribution of the component particles, such homogenized composite
 materials (HCMs) can exhibit optical
 properties not exhibited at all by their component materials, or at least not exhibited
 to the same extent by their components. Indeed, through judicious design,  HCMs arising from commonplace
component materials may be conceptualized
  which support rather  exotic (and potentially-useful) optical
phenomenons, while their component materials do not. A prime example
is provided by HCMs which support plane-wave propagation with
negative phase velocity \c{ML_JAP,ML_FCM}. Another example~---~which
provides  the setting for the present study~---~is furnished by HCMs
that  support a singular form of optical propagation known as
 Voigt wave propagation \c{ML_Voigt1}.

In order to  describe what constitutes a Voigt wave, it is helpful
to first consider the nonsingular case of optical propagation in a
linear, homogeneous, anisotropic, dielectric material. Generally two
independent plane waves, with orthogonal polarizations and different
phase velocities, can propagate in a given direction \c{BW}.
However, as first reported by Voigt in 1902 \c{Voigt} and later
described by others \c{Panch,Khap,Fedorov,Agranovich,Berry_Dennis},
in certain dissipative  biaxial crystals there are particular
directions along which these two waves coalesce to form a single
plane wave. A prominent feature of this coalescent Voigt wave is
that its amplitude has a linear dependence upon propagation
direction. Bianisotropic materials offer greater scope for Voigt
waves \c{Berry}, but herein  our attention is restricted to the
simpler case biaxial dielectric materials.

Previously, the standard Bruggeman homogenization formalism was used
to establish that certain dissipative biaxial HCMs can support Voigt
wave propagation \c{ML_Voigt1}. A follow-up study based on the
second-order strong-permittivity-fluctuation theory~---~which
represents a higher-order formulation of the standard Bruggeman
formalism wherein two-point statistical correlations between
particles of the component materials are taken into account
\c{Zhuck}~---~emphasized the importance of correlation length for
Voigt wave propagation \c{ML_Voigt2}. Both of these earlier studies
concerned HCMs arising from two uniaxial dielectric component
materials (which cannot themselves support Voigt wave propagation).
However, a biaxial dielectric HCM can also arise from isotropic
dielectric components in instances where the shapes of the component
material particles are non-spherical. For example, if each of the
two component materials comprises an assembly of oriented spheroidal
particles then the corresponding HCM will be biaxial, in general
\c{MW_biax2}. This is the scenario explored here.
 We implement
   an extended version of the
standard Bruggeman formalism \c{Goncharenko}, which utilizes a
recently-developed extended depolarization dyadic formalism
\c{M_WRM} in order to  take into account the nonzero size of
component particles.
 This approach
allows us to investigate the influence of the non-electromagnetic
attributes of the component materials~---~those being the
orientation, shape and size of the component particles as well as
the volume fraction of the component materials~---~upon the
propagation of Voigt waves.

In the notation adopted, vectors are represented in boldface, with
the $\hat{}$ symbol denoting a unit vector. Thus, the unit Cartesian
vectors are written as $\hat{\#x}$, $\hat{\#y}$ and $\hat{\#z}$.
  Double underlining with normal typeface signifies a  3$\times$3 dyadic; the identity 3$\times$3 dyadic
is $\=I = \hat{\#x} \, \hat{\#x} + \hat{\#y} \, \hat{\#y} +
\hat{\#z} \, \hat{\#z}$; and the superscript $T$ denotes the dyadic
transpose. Double underlining with blackboard bold typeface
signifies a 6$\times$6 dyadic.
  The
permittivity and permeability of free space are written as $\epso$
and $\muo$, respectively. The free-space wavenumber is $\ko = \omega
\sqrt{\epso \muo}$, with $\omega$ being the angular frequency.

\section{Homogenization formalism}

\subsection{Component materials}

We consider optical propagation in an HCM derived from two component
materials, labelled $a$ and $b$. The component materials are both
taken to be  isotropic dielectric materials with permittivity
dyadics $\=\eps_{\,a} = \epso \eps_a \=I$ and $\=\eps_{\,b} = \epso
\eps_b \=I$.  The volume fraction of material $a$ is $f_a$ while
that of material $b$ is $f_b = 1 - f_a$.

Each component material comprises a randomly-distributed assembly of
spheroidal particles. All material $a$ particles have the same
orientation and all material $b$ particles have the same
orientation, but these two orientations are generally different. For
simplicity,  both material $a$ and $b$ particles are assumed to have
the same shape.
 The surfaces of the
component particles, relative to their centres,  are prescribed by
the vector
\begin{equation}
\#r_\ell = \eta_\ell \, \=U_{\,\ell} \. \hat{\#r}, \qquad (\ell = a,
b).
\end{equation}
Here, $\hat{\#r}$ is the radial  vector prescribing the surface of
the unit sphere, the  real-symmetric surface dyadic $\=U_{\,\ell}$
maps the spherical surface  onto a spheroidal one, and $\eta_\ell >
0$ is linear measure of particle size. In conformity with the
homogenization regime, $\eta_\ell$ must be much smaller than the
wavelengths involved, but~---~unlike in conventional approaches
taken in homogenization studies \c{EAB}~---~we shall  not insist
that $\eta_\ell$ is vanishingly small. Furthermore, let us assume
that $\eta_a \equiv \eta_b$, and henceforth simply write $\eta$ in
lieu of $\eta_\ell$.

The symmetry axis of the spheroidal particles comprising material
$a$ is
 taken to lie in the $xy$ plane at an angle $\varphi$ to the $x$ axis.
 Thus, the surface dyadic for material $a$  may be expressed as
\begin{equation}
\=U_{\,a} = \frac{1}{\sqrt[3]{U_x U^2 }} \, \=R_{\,z} (\varphi) \.
\les U_x \hat{\#x} \, \hat{\#x} + U \le \hat{\#y} \, \hat{\#y} +
\hat{\#z} \, \hat{\#z} \ri  \ris \. \=R^T_{\,z} (\varphi), \qquad
(U, U_x > 0),
\end{equation}
where the orthogonal rotation dyadic
\begin{equation}
\=R_{\,z} (\varphi) =
 \cos \varphi
\le  \, \hat{\#x} \, \hat{\#x} + \hat{\#y} \, \hat{\#y} \, \ri +
 \sin \varphi
\le  \, \hat{\#x} \, \hat{\#y} - \hat{\#y} \, \hat{\#x} \, \ri +
\hat{\#z} \, \hat{\#z}.
\end{equation}
Without loss of generality, the material $b$ particles are assumed
to be aligned with the $x$ axis. Thus, the surface dyadic for
material $b$  may be expressed as
\begin{equation}
\=U_{\,b} = \frac{1}{\sqrt[3]{U_x U^2 }} \les U_x \hat{\#x} \,
\hat{\#x} + U \le \hat{\#y} \, \hat{\#y} + \hat{\#z} \, \hat{\#z}
\ri  \ris, \qquad (U, U_x > 0) .
\end{equation}

\subsection{Homogenized composite material}

As the symmetry axes of the spheroidal particles comprising
components $a$ and  $b$  are not generally aligned, the resulting
HCM is a biaxial dielectric material with a symmetric permittivity
dyadic of the form
\begin{equation}
\=\eps_{\,HCM} = \epso \les \, \eps_x   \, \hat{\#x} \, \hat{\#x} +
\eps_y \hat{\#y} \, \hat{\#y}  +
 \eps_t \le  \hat{\#x} \, \hat{\#y} + \hat{\#y} \, \hat{\#x} \, \ri + \eps_z \hat{\#z}
\, \hat{\#z} \, \ris.
\end{equation}
We implement an extended version of the Bruggeman homogenization
formalism in order to estimate $\=\eps_{\,HCM}$. Accordingly,
$\=\eps_{\,HCM}$ is extracted from the nonlinear dyadic equation
\c{M_Electromagnetics}
\begin{equation}
f_a \lec \le \=\eps_{\,a} -  \=\eps_{\,HCM} \ri \. \les \=I +
\=D_{\,a} \. \le \=\eps_{\,a} -  \=\eps_{\,HCM} \ri \ris^{-1} \ric +
f_b \lec \le \=\eps_{\,b} -  \=\eps_{\,HCM} \ri \. \les \=I +
\=D_{\,b} \. \le \=\eps_{\,b} -  \=\eps_{\,HCM} \ri \ris^{-1} \ric =
0,
 \end{equation}
  using standard numerical techniques, such as the Jacobi method \c{Buchanan}.
The depolarization dyadics $\=D_{\,a,b}$ herein may be regarded as
sums of two terms; that is,
\begin{equation}
\=D_{\,\ell} = \=D^{0}_{\,\ell} +  \=D^{+}_{\,\ell}, \qquad (\ell =
a, b).
\end{equation}
The term $\=D^{0}_{\,\ell}$ represents the depolarization
contribution arising from a vanishingly small particle described by
the surface dyadic  $\=U_{\,\ell}$, as given by the double integral
\c{M97,MW97}
\begin{equation} \l{D_o}
\=D^{0}_{\,\ell} = \frac{1}{4 \pi}\, \int^{2 \pi}_{\phi = 0}
\int^{\pi}_{\theta = 0} \frac{ \le \=U^{-1}_{\, \ell} \. \hat{\#q}
\ri \,  \le \=U^{-1}_{\, \ell} \. \hat{\#q} \ri \; \sin \theta}{\le
\=U^{-1}_{\, \ell} \. \hat{\#q} \ri \. \=\eps_{\, HCM} \. \le
\=U^{-1}_{\, \ell} \. \hat{\#q} \ri}\, d \theta\, d \phi \, , \qquad
(\ell = a, b),
\end{equation}
wherein the unit vector $\hat{\#q} = \sin \theta \cos \phi \,
\hat{\#x} + \sin \theta \sin \phi \, \hat{\#y} + \cos \theta \,
\hat{\#z}$. The depolarization contribution which derives from the
nonzero size of the component particles is provided by the
 term $\=D^{+}_{\,\ell}$. It is most conveniently expressed
 in terms of  elements of the  6$\times$6 dyadic $\underline{\underline{\mathbb{D}}}^{+}_{\,\ell}$, per
\begin{equation}
\les \, \=D^{+}_{\,\ell} \, \ris_{m n} = \les \,
\underline{\underline{\mathbb{D}}}^{+}_{\,\ell} \, \ris_{m n},
\qquad \le m, n \in \lec 1, 2, 3 \ric \ri,
\end{equation}
with \c{M_WRM}
\begin{eqnarray} \l{D_plus}
\underline{\underline{\mathbb{D}}}^{+}_{\,\ell} &=&
\frac{\omega^4}{4 \pi \muo } \int^{2 \pi}_{\phi = 0}
\int^{\pi}_{\theta = 0} \frac{\sin \theta}{\les \le \=U^{-1}_{\,
\ell} \. \hat{\#q} \ri \. \=\eps_{\, HCM} \. \le \=U^{-1}_{\, \ell}
\. \hat{\#q} \ri \ris \le \=U^{-1}_{\, \ell} \. \hat{\#q} \ri \. \le
\=U^{-1}_{\, \ell} \. \hat{\#q} \ri } \nonumber \\ && \times \Bigg[
\frac{1}{ \kappa_+ - \kappa_-  }  \Bigg( \frac{\exp \le i \eta q \ri
}{2 q^2} \le 1 - i \eta q\ri
 \Big\{ \,
 \mbox{det} \les \underline{\underline{\mathbb{A}}} (\=U^{-1}_{\, \ell} \.\#q ) \ris \, \underline{\underline{\mathbb{G}}}^{+} (\=U^{-1}_{\,
\ell} \.\#q) \nonumber \\
&&
  +  \mbox{det} \les \underline{\underline{\mathbb{A}}} (-\=U^{-1}_{\, \ell} \.\#q ) \ris  \,  \underline{\underline{\mathbb{G}}}^{+} (-\=U^{-1}\.\#q
) \Big\} \Bigg)^{q=\sqrt{ \kappa_+ }}_{q=\sqrt{ \kappa_- }} + \frac{
 \mbox{det} \les \underline{\underline{\mathbb{A}}} (\#0 ) \ris} {\kappa_+  \, \kappa_- }\,
\underline{\underline{\mathbb{G}}}^{+} (\#0)
 \Bigg] \; d \theta \; d \phi, \qquad (\ell = a,b). \nonumber \\ &&
\end{eqnarray}
Herein  $\kappa_\pm $ are the $q^2$ roots of $\mbox{det} \les
\underline{\underline{\mathbb{A}}}(\=U^{-1}\.\#q) \ris = 0$, the
vector $\#q = q \, \hat{\#q}$, while the 6$\times$6 dyadics
\begin{equation}
\underline{\underline{\mathbb{A}}} (\#p)  = \les \begin{array}{cc}
\=\eps_{\,HCM} & \le \#p / \omega \ri \times \=I \vspace{2mm}  \\
-\le \#p / \omega \ri \times \=I  & \muo \, \=I
\end{array}
 \ris
\end{equation}
and
\begin{equation}
\underline{\underline{\mathbb{G}}}^{+} (\#p) =
\underline{\underline{\mathbb{A}}}^{-1} (\#p) - \lim_{| \#p | \to
\infty} \underline{\underline{\mathbb{A}}}^{-1} (\#p) .
\end{equation}
Analytical evaluations of the integrals in eqs.~\r{D_o} and
\r{D_plus} are available for relatively simple anisotropic HCMs
\c{W98,M_JNP}, but for general biaxial HCMs numerical methods are
needed to evaluate these integrals.

\section{Voigt wave propagation}

Let us turn now to  the possibility of Voigt wave propagation in the
HCM. All propagation directions relative to the symmetry axes of the
HCM should be considered. It is expedient to do so indirectly, by
investigating  Voigt wave propagation along the $z$ axis for all
possible orientations of the HCM. Thus, we introduce the  HCM
permittivity dyadic in the rotated coordinate frame
\begin{eqnarray}
\={\tilde{\eps}}_{\,HCM} (\alpha, \beta, \gamma ) &=&
 \=R_{\,z}(\gamma)\.\=R_{\,y}(\beta)\.\=R_{\,z}(\alpha)\.\=\eps_{\,HCM}\.
 \=R^T_{\,z}(\alpha)\.\=R^T_{\,y}(\beta)\.\=R^T_{\,z}(\gamma) \l{HCMrot2} \\
 &=& \eps_{11} \,
\hat{\#x} \, \hat{\#x} + \eps_{22} \, \hat{\#y} \,  \hat{\#y} +
\eps_{33} \, \hat{\#z} \,  \hat{\#z} +  \eps_{12} \, \le \hat{\#x}
\, \hat{\#y} + \hat{\#y} \,  \hat{\#x} \ri \nonumber \\ &&  +
\eps_{13} \, \le \hat{\#x} \, \hat{\#z} + \hat{\#z} \,  \hat{\#x}
\ri + \eps_{23} \, \le \hat{\#y} \, \hat{\#z} + \hat{\#z} \,
\hat{\#y} \ri,
 \l{HCMrot1}
\end{eqnarray}
where the orthogonal rotation dyadic
\begin{equation}
\=R_{\,y} (\beta)= \cos \beta \le  \, \hat{\#x} \, \hat{\#x} +
\hat{\#z} \, \hat{\#z} \, \ri +
 \sin \beta
\le  \, \hat{\#z} \, \hat{\#x} - \hat{\#x} \, \hat{\#z} \, \ri +
\hat{\#y} \, \hat{\#y},
\end{equation}
and  $\alpha$, $\beta$ and $\gamma$ are the three Euler angles
\c{Arfken}.

In order for Voigt waves to propagate along the $z$ axis of the
biaxial dielectric material described by the permittivity dyadic
\r{HCMrot2}, the following two conditions must be satisfied
\c{GL01}:
\begin{itemize}  \item[(i)] $Y (\alpha, \beta, \gamma ) = 0$, and
\item[(ii)] $W (\alpha, \beta, \gamma ) \neq 0$.
\end{itemize}
where the scalars
\begin{eqnarray}
Y(\alpha, \beta, \gamma ) &=& \eps^4_{13} + \eps^4_{23} -2
\eps_{23}\eps_{33}
 \les \, 2 \eps_{12}
\eps_{13} - \le \, \eps_{11} - \eps_{22}\, \ri \eps_{23}\,\ris +
\les \le \, \eps_{11}-\eps_{22}\,\ri^2 + 4 \eps^2_{12}\,\ris \,
\eps^2_{33} \nonumber \\ && + 2 \eps_{13} \lec \, \eps^2_{23}
\eps_{13} - \les \, 2 \eps_{12}\eps_{23} + \le \, \eps_{11} -
\eps_{22} \, \ri \, \eps_{13}\,\ris \eps_{33}\,\ric
\end{eqnarray}
and
\begin{equation}
W(\alpha, \beta, \gamma ) = \eps_{12} \eps_{33} - \eps_{13}
\eps_{23}\,.
\end{equation}

Let us note that the conditions (i) and (ii) cannot be satisfied by
isotropic or  uniaxial  dielectric materials.

\section{Numerical studies}

\subsection{Preliminaries}

We now investigate the scope for Voigt wave propagation in the
biaxial HCM, in terms of the shape, size and orientation of the
component material particles, as well as the  volume fraction of the
component materials,
 by means of representative numerical
calculations. Two stages are involved:
 first, $\=\eps_{\,HCM}$ is estimated using the extended Bruggeman formalism; and second,
 the quantities
$Y(\alpha, \beta, \gamma )$ and $W(\alpha, \beta, \gamma )$ are
calculated as functions of the Euler angles. More specifically, the
 angular coordinates $(\alpha, \beta, \gamma )$ of the zeros of $|Y|$,
and the corresponding values of $| W |$  are computed. Notice that
the angular coordinate $\gamma$  may be  eliminated from our
investigation because propagation parallel to the $z$ axis (of the
rotated coordinate system) is independent of rotation about that
axis.

In the following, we chose   $\eps_a = 1.5$ and $\eps_b = 12$ as the
relative permittivities of the  component materials $a$ and $b$.
Since $\eps_{a,b} \in \mathbb{R}$, the component materials are
nondissipative. However, under the extended Bruggeman homogenization
formalism with $\eta > 0$, the relative permittivity parameters of
the HCM are complex-valued. The imaginary parts of the HCM's
relative permittivity parameters are indicative of losses due to
scattering from the macroscopic coherent field \c{Kranendonk}. This
is a general feature of higher-order approaches to homogenization,
as  occurs in a similar fashion with the
strong-permittivity-fluctuation theory \c{TKN}, for example.

We take the spheroid parameters $U_x = 1 + \rho $ and $U = 1- \le
\rho / 18 \ri$, and consider the range $0 < \rho < 9$. Thus, the
component particles become increasingly elongated as the
eccentricity parameter $\rho$ increases from zero, whereas in the
limit $\rho \to 0$ the particle shape becomes spherical.

\subsection{HCM constitutive parameters}

The extended Bruggeman estimates of the HCM's relative permittivity
parameters $\eps_{x,y,z,t}$ are plotted as functions of spheroid
orientation angle $\varphi$ and volume fraction $f_a$ in
Fig.~\ref{fig1}.  Here, we fixed the relative size parameter $ \eta
= 0.2/\ko$ and the eccentricity parameter $\rho = 9$. We see that
the real parts of $\eps_{x,y,z}$ decay almost linearly as $f_a$
increases from 0 to 1, but $\mbox{Re} \lec \eps_{x,y,z} \ric$
 are largely
insensitive to variations in $\varphi$. While the real part of
$\eps_t$ is nonexistent for $\varphi = 0^\circ$ and $90^\circ$, and
also in the limits $f_a \to 0$ and $1$,  away from these boundary
values we have $\mbox{Re} \lec \eps_{t} \ric < 0$ with a local
minimum occurring at approximately $\varphi = 45^\circ$ and $f_a =
0.26$. The imaginary parts of $\eps_{x,y,z}$ attain their largest
values in the vicinity of $f_a = 0.26$, and vanish at $f_a = 0$ and
1. As $\varphi$ increases from zero, $\mbox{Im} \lec \eps_{x} \ric $
generally increases, $\mbox{Im} \lec \eps_{y} \ric $ generally
decreases, and $\mbox{Im} \lec \eps_{z} \ric $ is largely unchanged.
 The imaginary part of $\eps_t$ is null-valued along the boundaries
$\varphi = 0^\circ$ and $90^\circ$, and $f_a = 0$ and $1$, but away
from these boundary values  $\mbox{Im} \lec \eps_{t} \ric $ exhibits
a local maximum which coincides with the local minimum exhibited by
$\mbox{Re} \lec \eps_{t} \ric $. The following two limits, which
hold for arbitrary $f_a \in \le 0, 1 \ri$, are especially
noteworthy: (a) As $\varphi \to 0^\circ$, we find that $\eps_x \neq
\eps_y = \eps_z$; i.e., the HCM is uniaxial. Accordingly the HCM
cannot support Voigt wave propagation in this limit \c{GL01}. (b)
As $\varphi \to 90^\circ$, we find that $\eps_x$, $ \eps_y $ and
$\eps_z$ are all different while $\eps_t = 0$; i.e., the HCM's
biaxial structure is orthorhombic.

Next we turn to the dependency upon the component particle shape and
size. In Fig.~\ref{fig2}, $\eps_{x,y,z,t}$ are plotted against
relative  size parameter $\ko \eta$ and the eccentricity parameter
$\rho$. Here, we fixed the volume fraction $ f_a = 0.25$ and the
spheroid orientation angle $\varphi = 45^\circ$. The real parts of
$\eps_{x,y,z}$ generally increase as $\eta$ increases, whereas
$\mbox{Re} \lec \eps_{t} \ric$ is largely independent of $\eta$. As
$\rho$ increases,  $\mbox{Re} \lec \eps_x \ric$ generally increases
but $\mbox{Re} \lec \eps_{y,z,t} \ric$
  generally decrease.
The most conspicuous feature of the plots of $\mbox{Im} \lec
\eps_{x,y,z,t} \ric$ is that these quantities increase uniformly
from zero as $\eta$ increases from zero. Also, $\mbox{Im} \lec
\eps_{x,y,z} \ric$ are fairly insensitive to increasing $\rho$ but
$\mbox{Im} \lec \eps_{t} \ric$ increases markedly as $\rho$
increases. Let us note that in the limit $\rho \to 0$, we have
$\eps_x = \eps_y = \eps_z$ and $\eps_t = 0$; i.e., the HCM is an
isotropic dielectric material, regardless of the size parameter
$\eta$ (or the orientation angle $\varphi$ and volume fraction
$f_a$).

\subsection{Orientations for Voigt waves}

\begin{figure}[!h]
\centering \psfull \epsfig{file=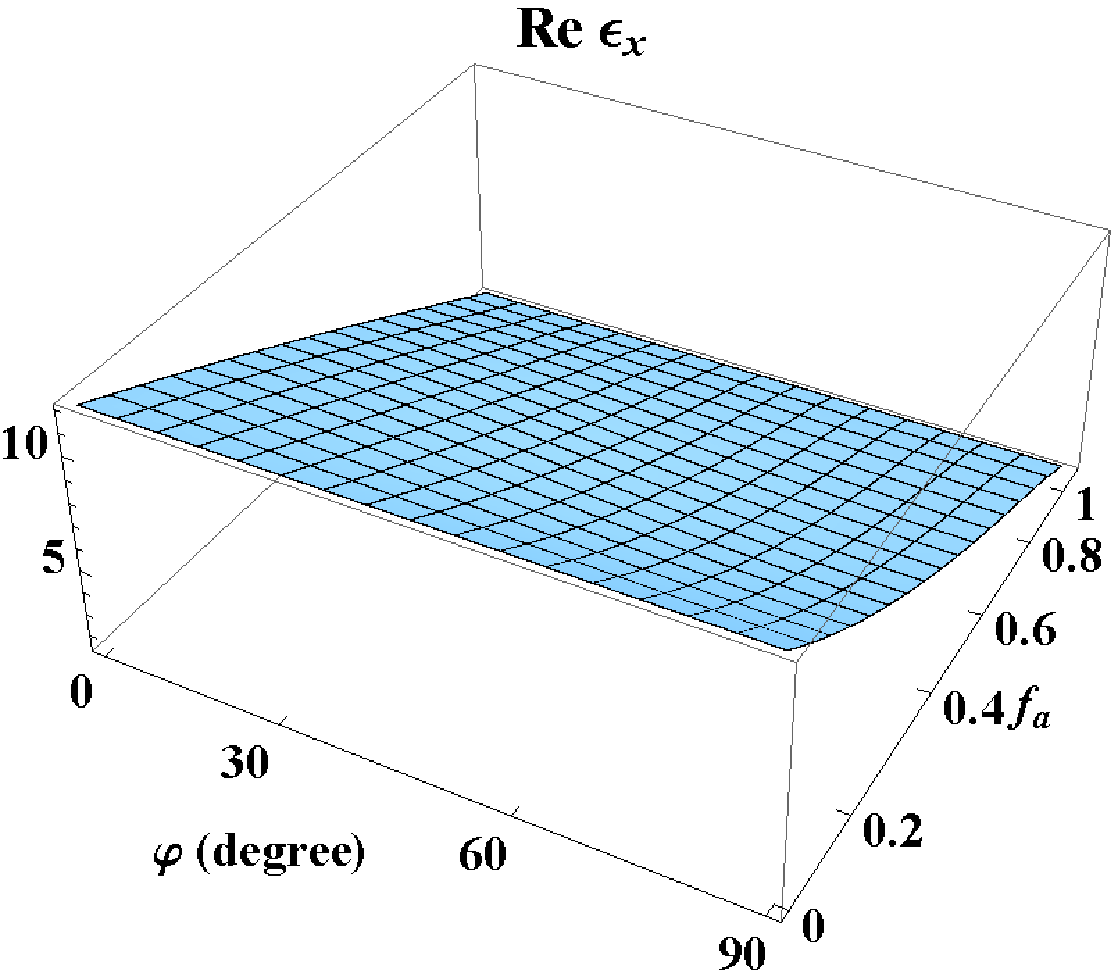,width=2.2in}
\hspace{20mm}
\epsfig{file=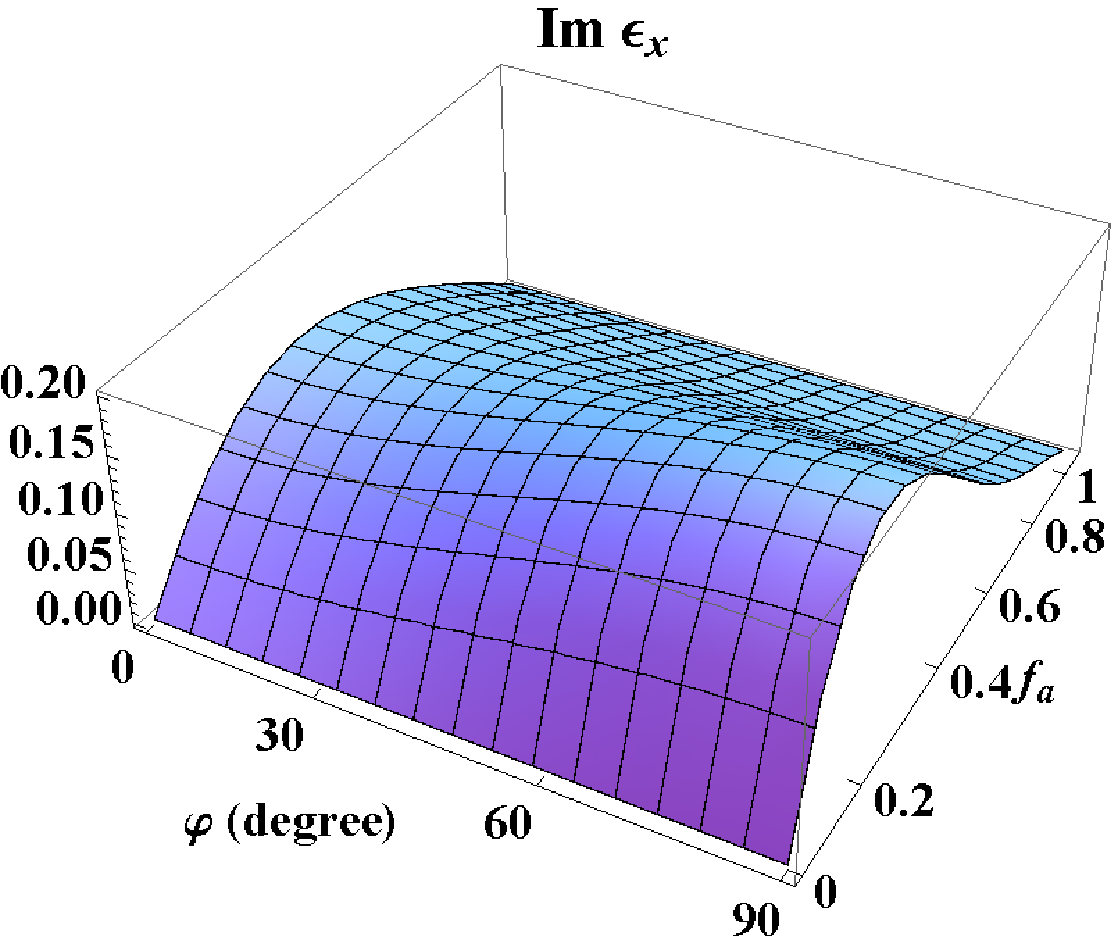,width=2.2in} \\
\epsfig{file=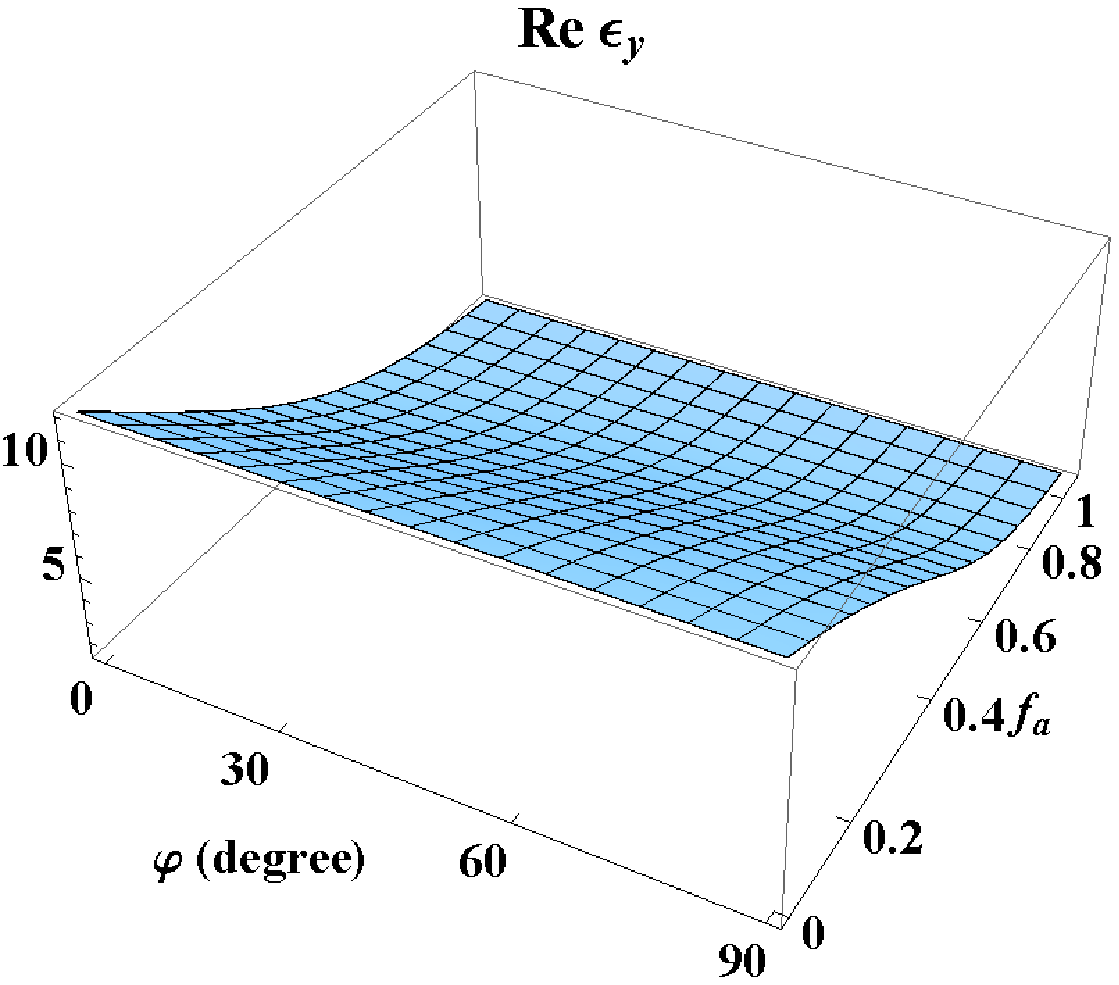,width=2.2in} \hspace{20mm}
\epsfig{file=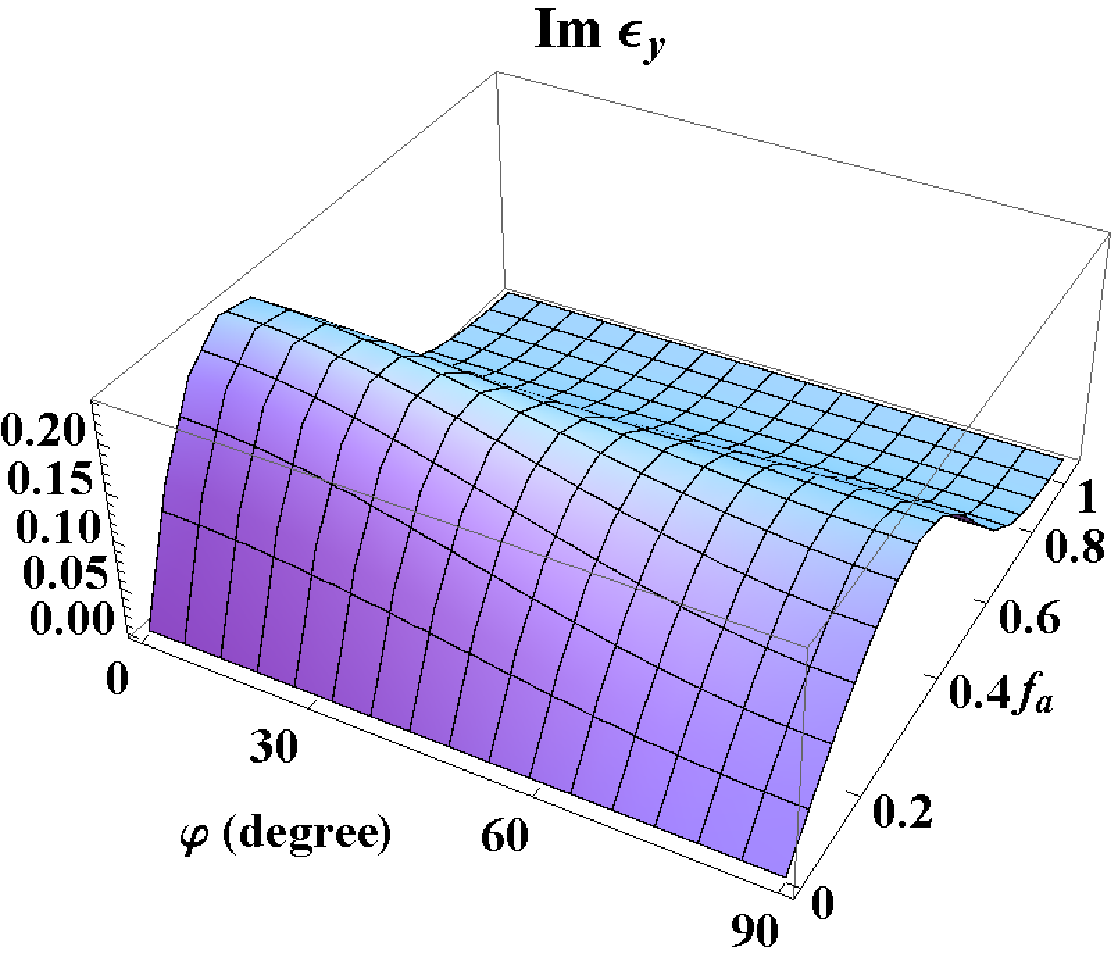,width=2.2in} \\
\epsfig{file=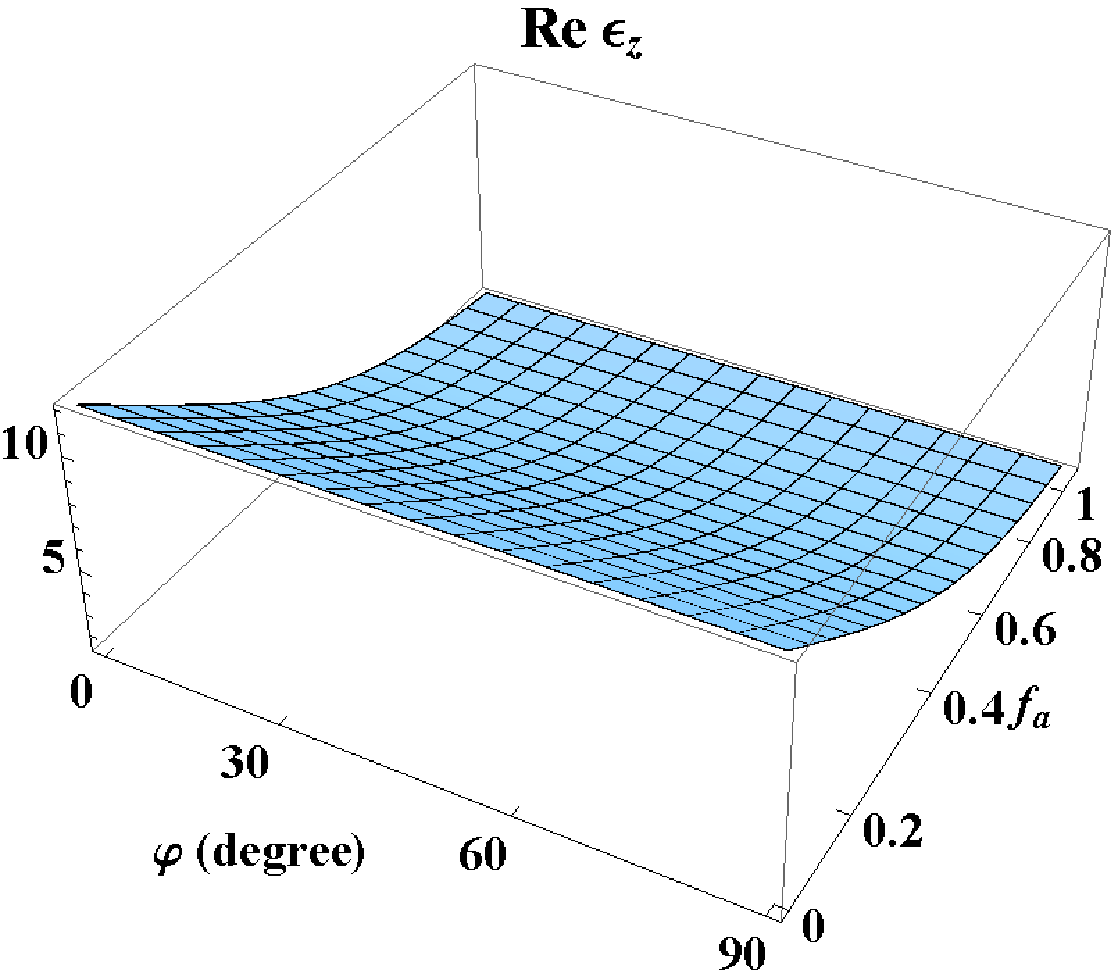,width=2.2in} \hspace{20mm}
\epsfig{file=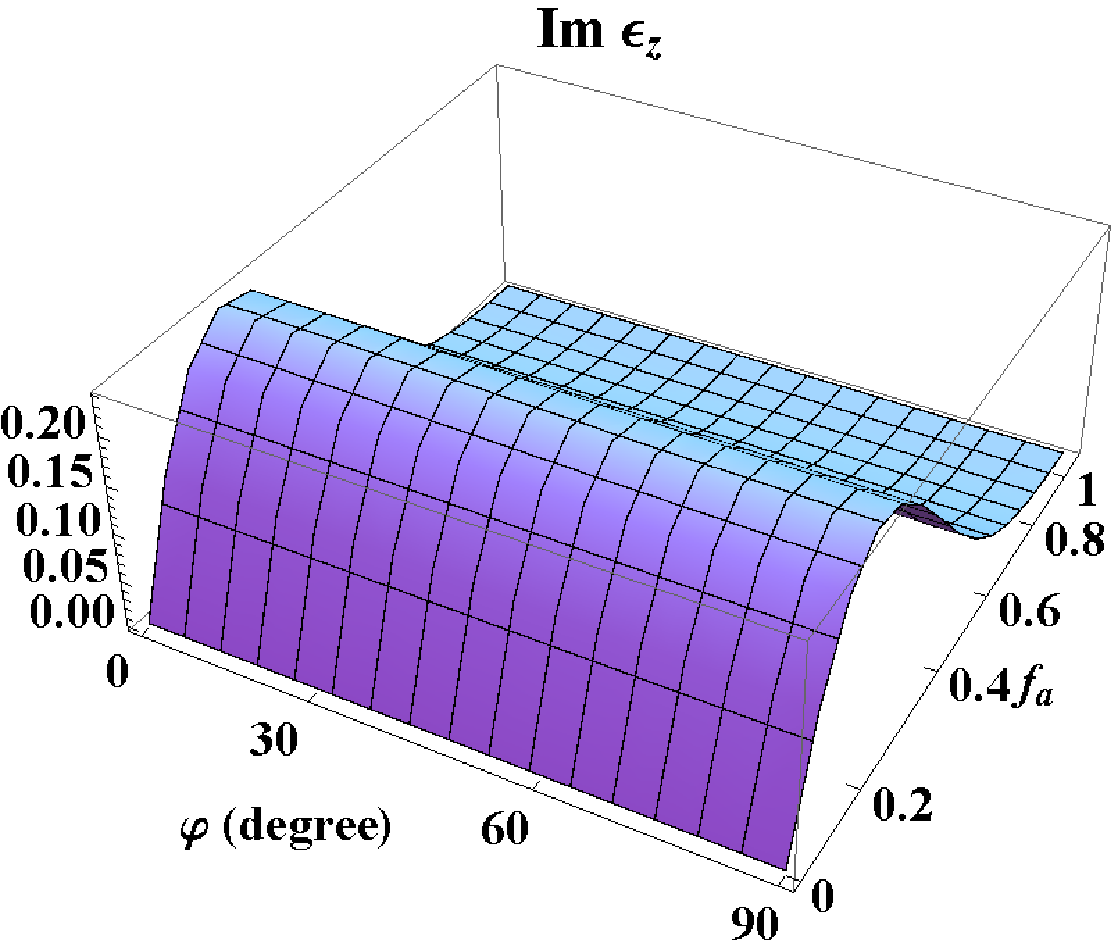,width=2.2in} \\
\epsfig{file=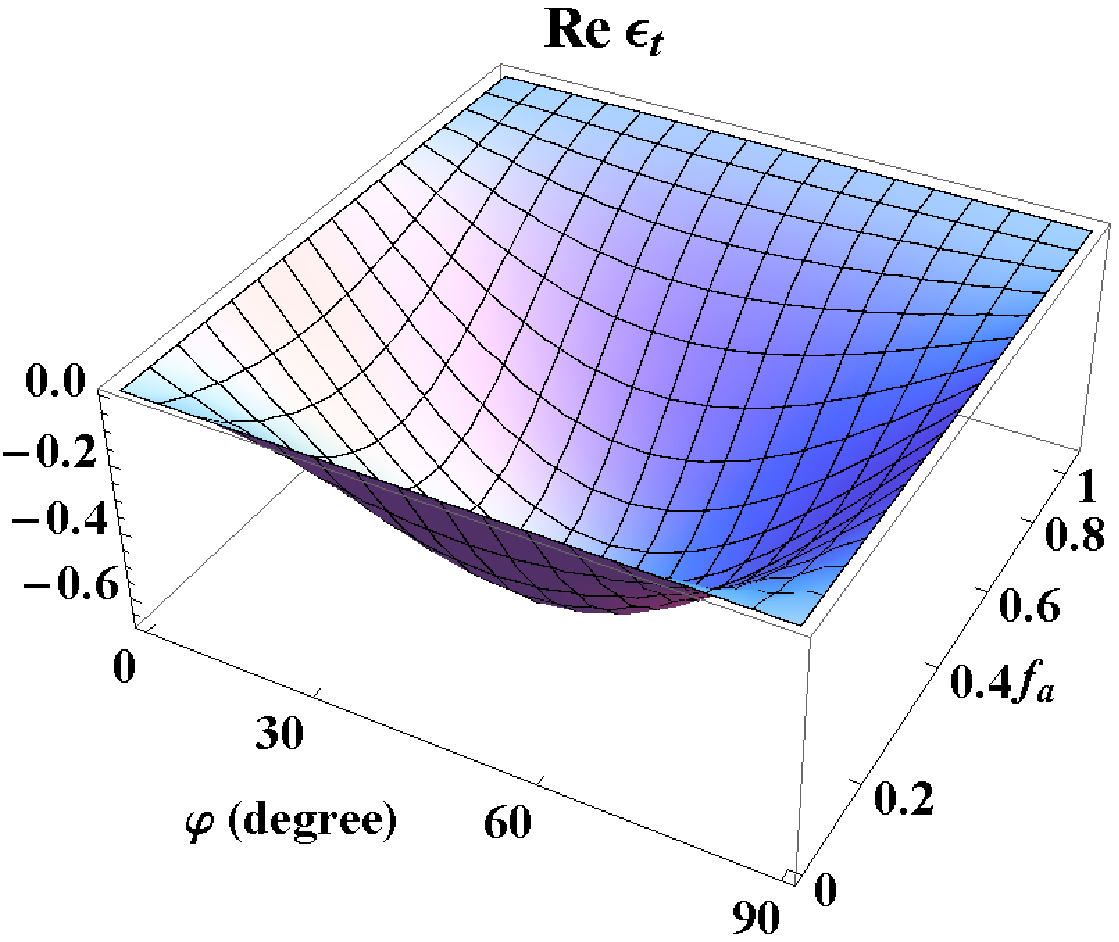,width=2.2in} \hspace{20mm}
\epsfig{file=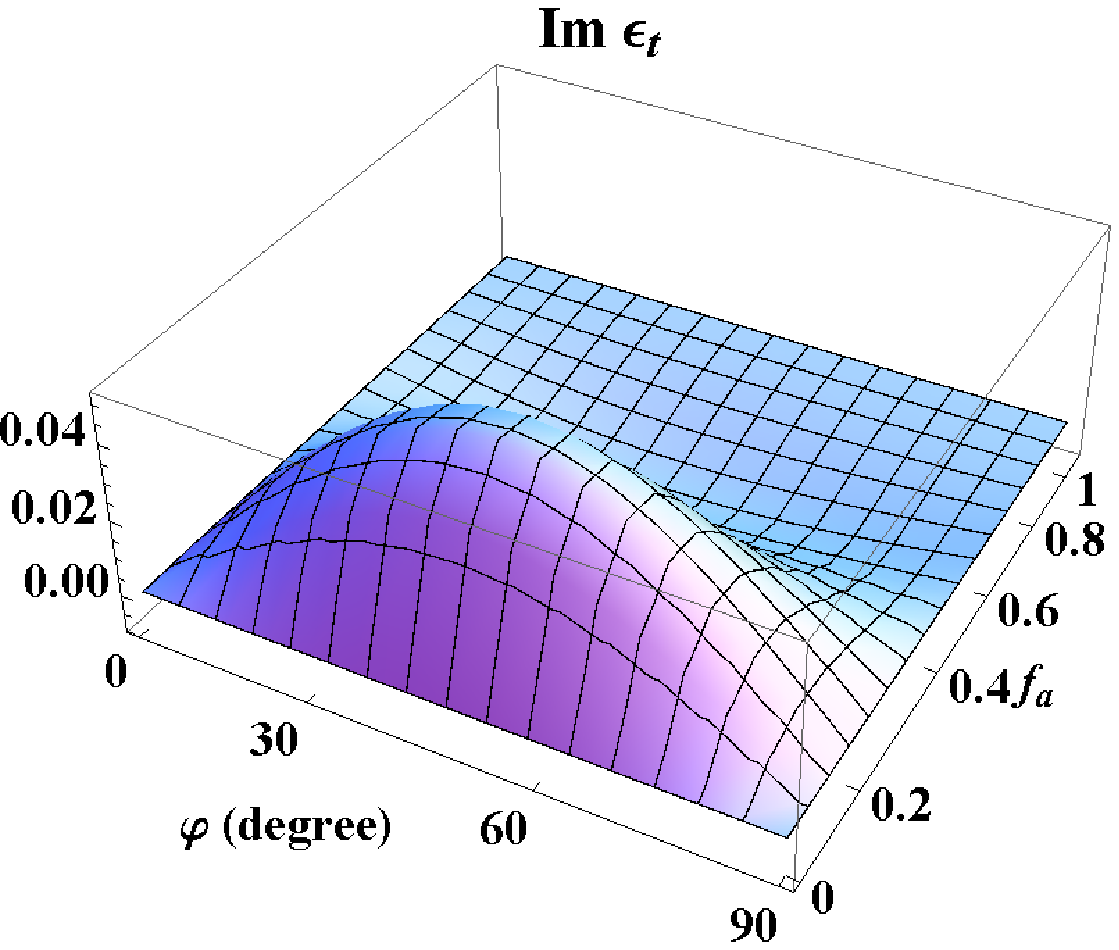,width=2.2in}
 \caption{The extended Bruggeman estimates of relative permittivity parameters of the HCM plotted versus volume fraction $f_a \in \le 0, 1 \ri$
 and spheroid orientation angle $\varphi \in \le 0^\circ, 90^\circ \ri$.
The relative size parameter $\ko \eta = 0.2$ and the eccentricity
parameter $\rho = 9$. } \label{fig1}
\end{figure}

\begin{figure}[!h]
\centering \psfull \epsfig{file=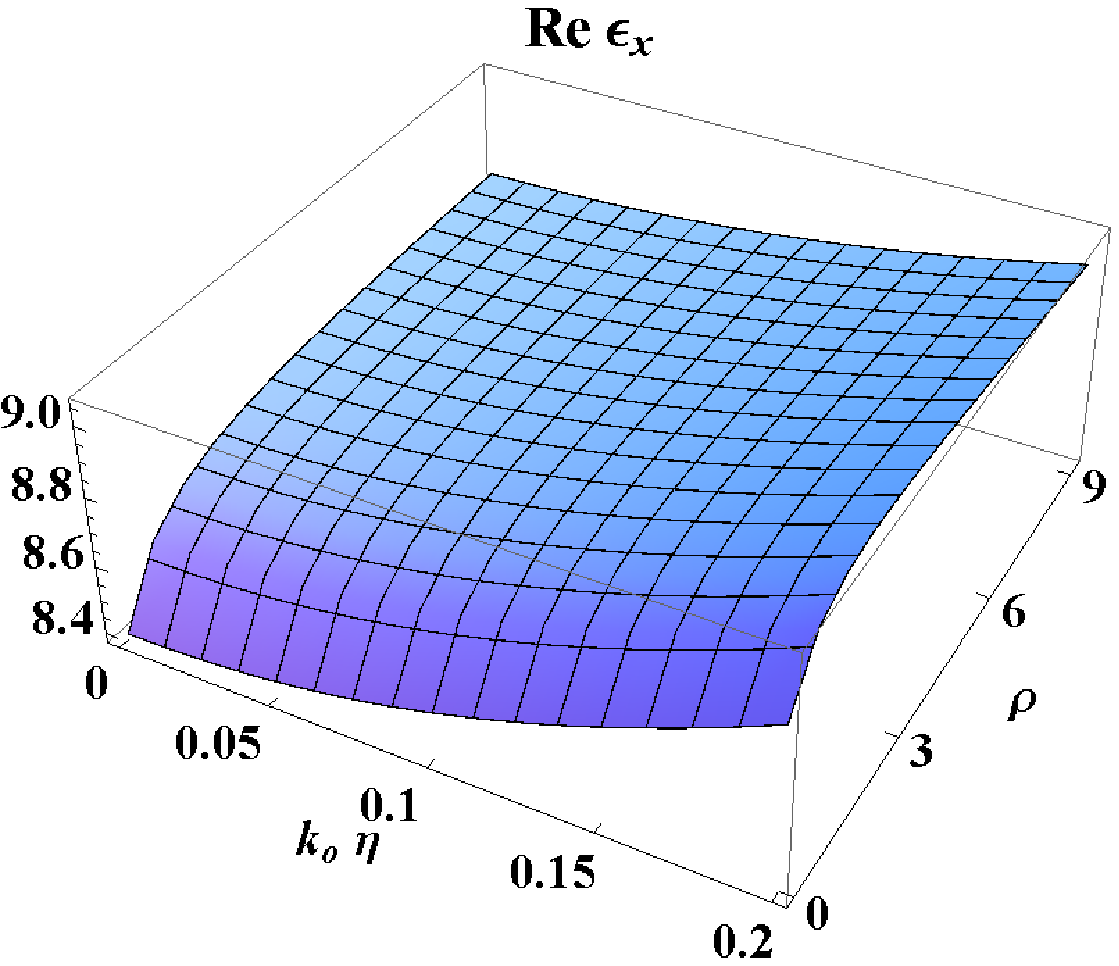,width=2.2in}
\hspace{20mm}
\epsfig{file=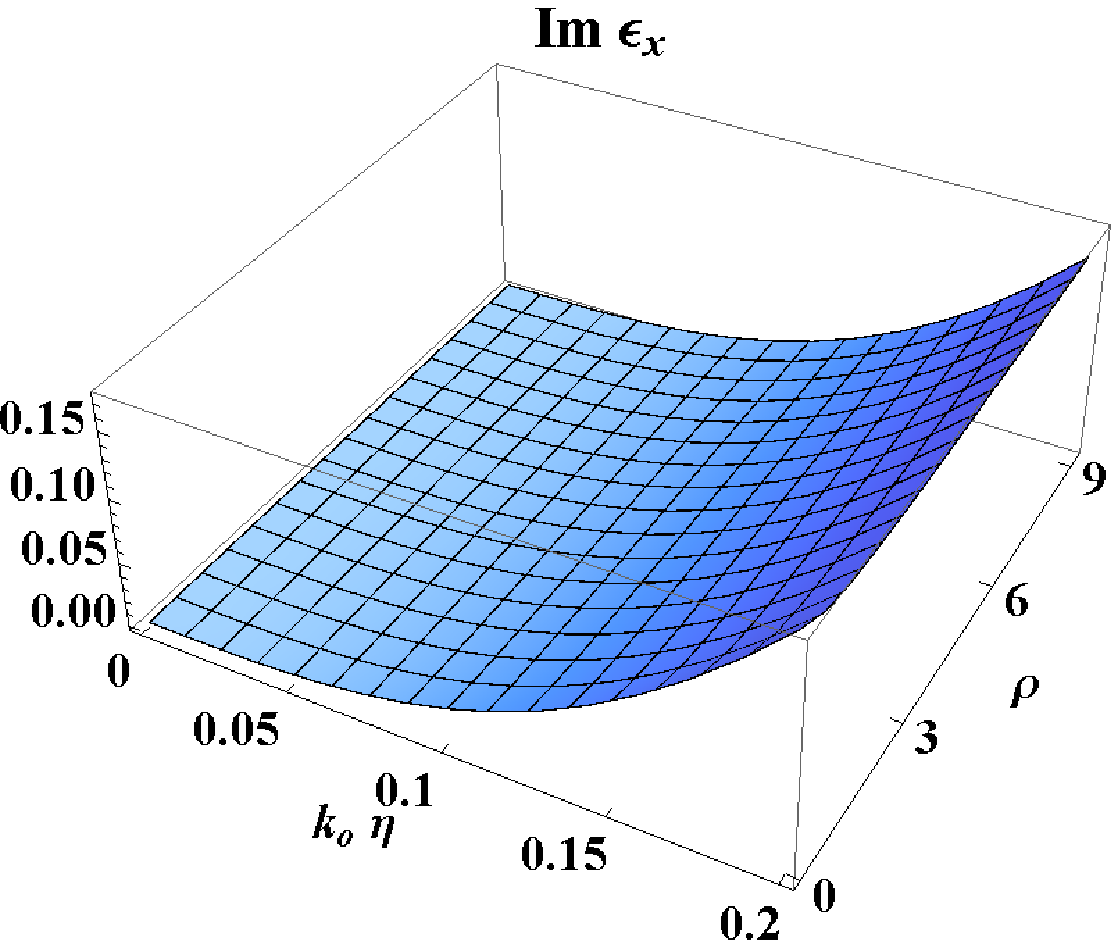,width=2.2in} \\
\epsfig{file=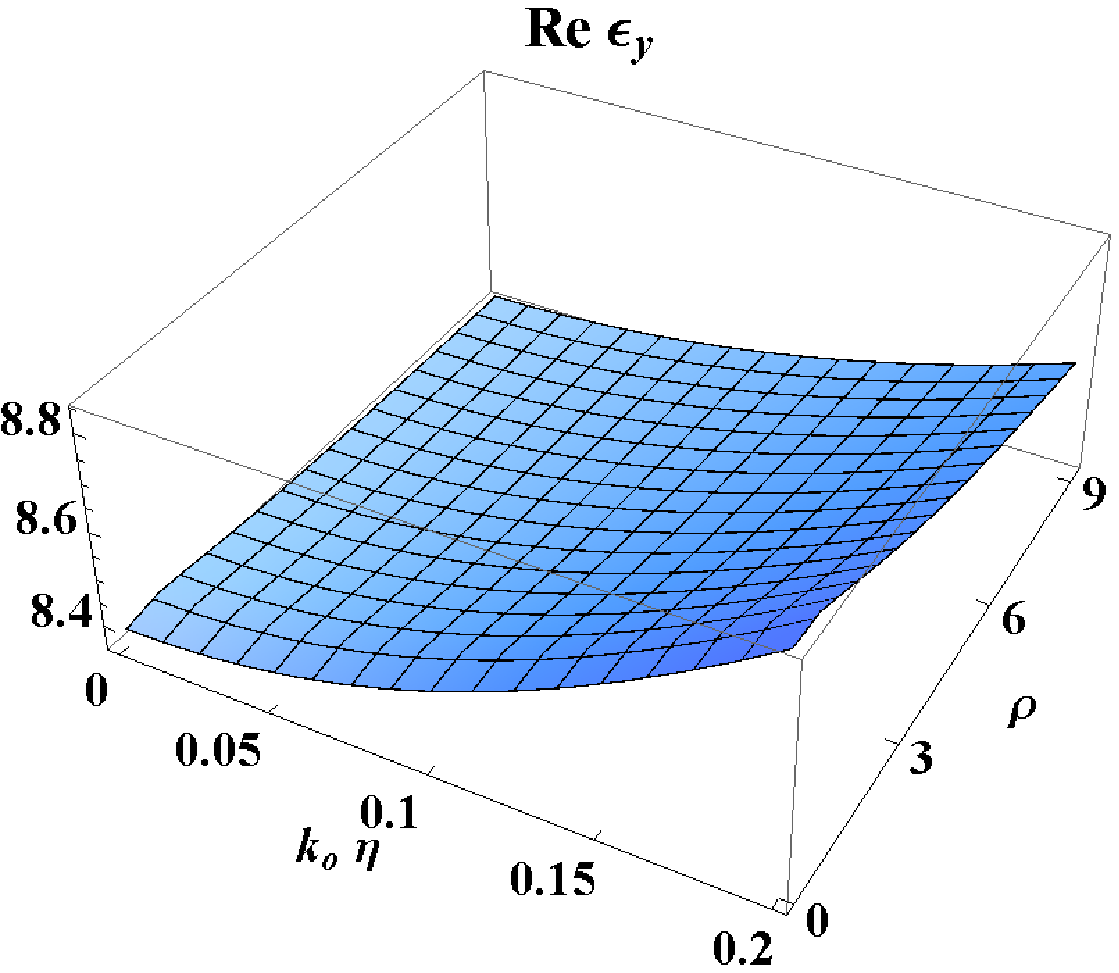,width=2.2in} \hspace{20mm}
\epsfig{file=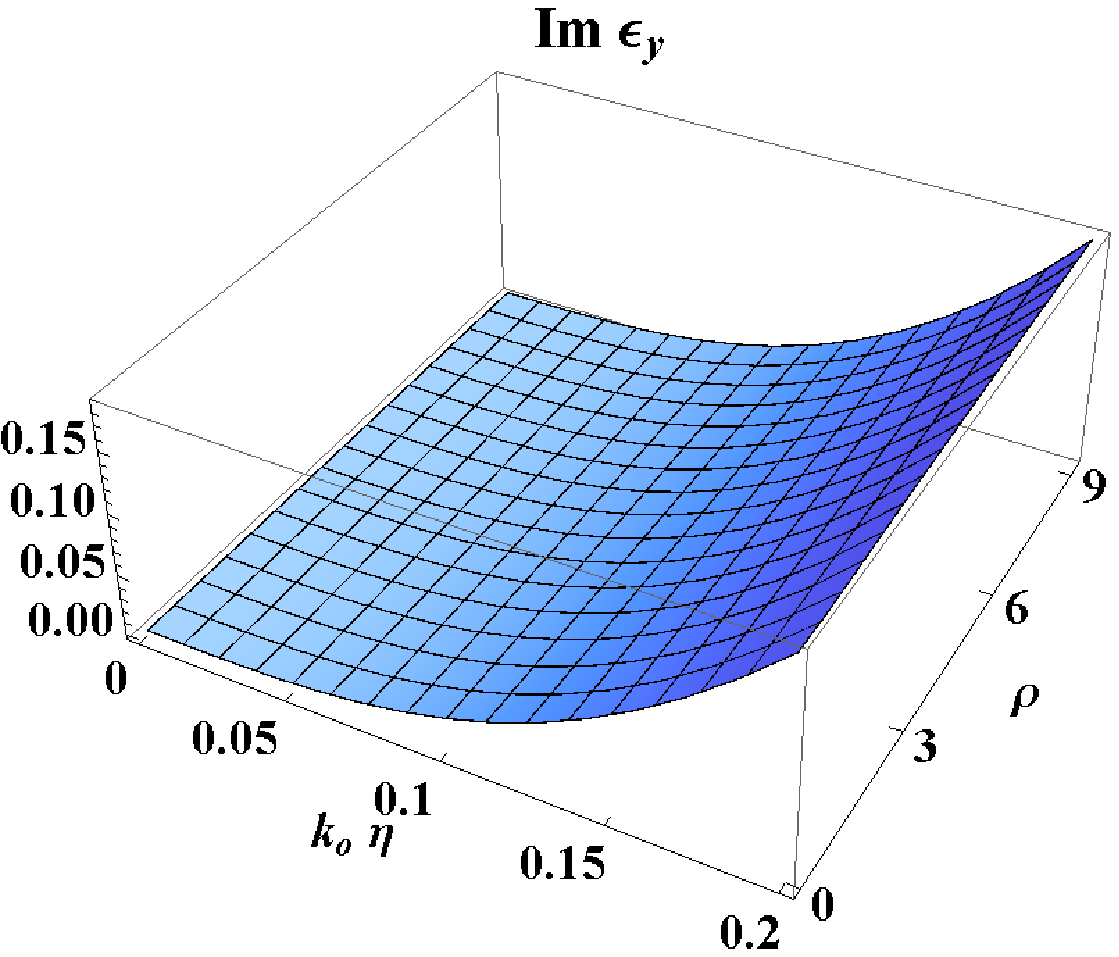,width=2.2in} \\
\epsfig{file=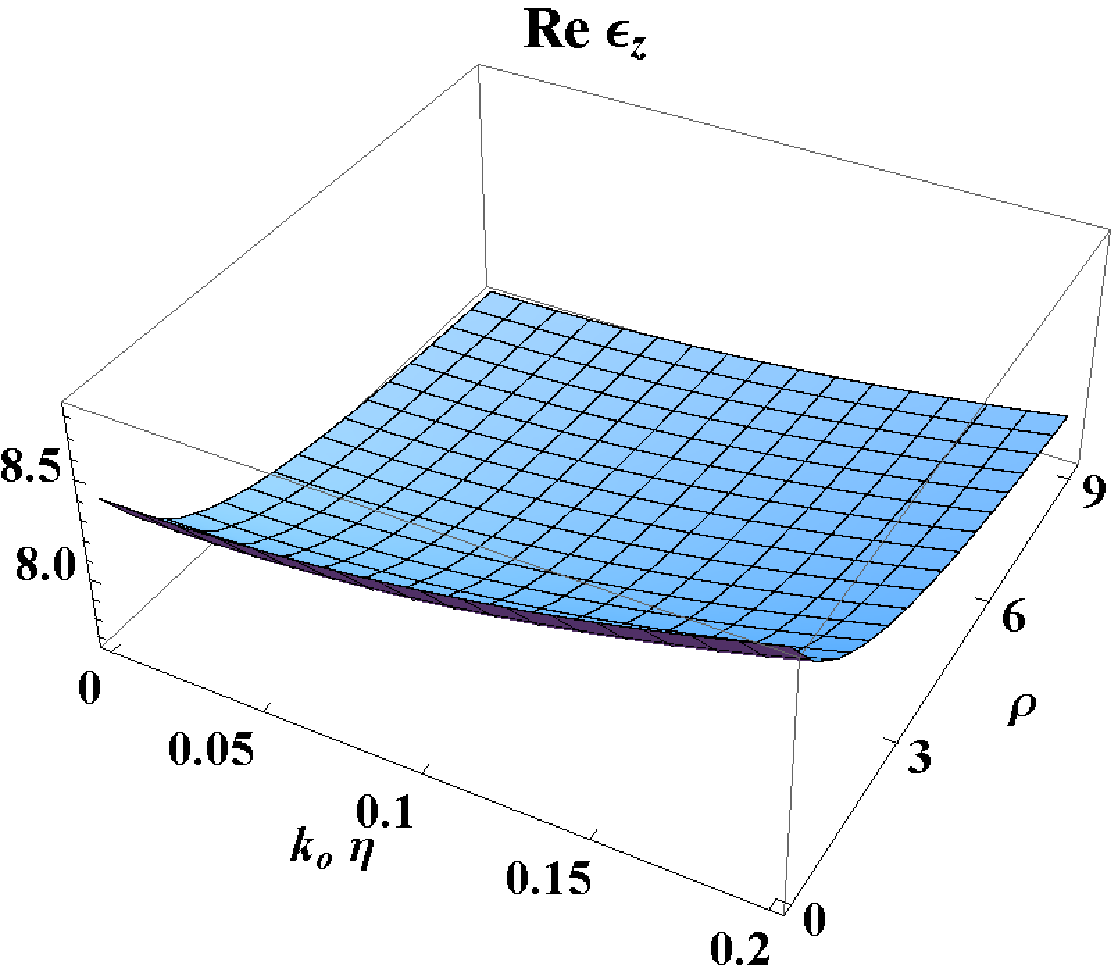,width=2.2in} \hspace{20mm}
\epsfig{file=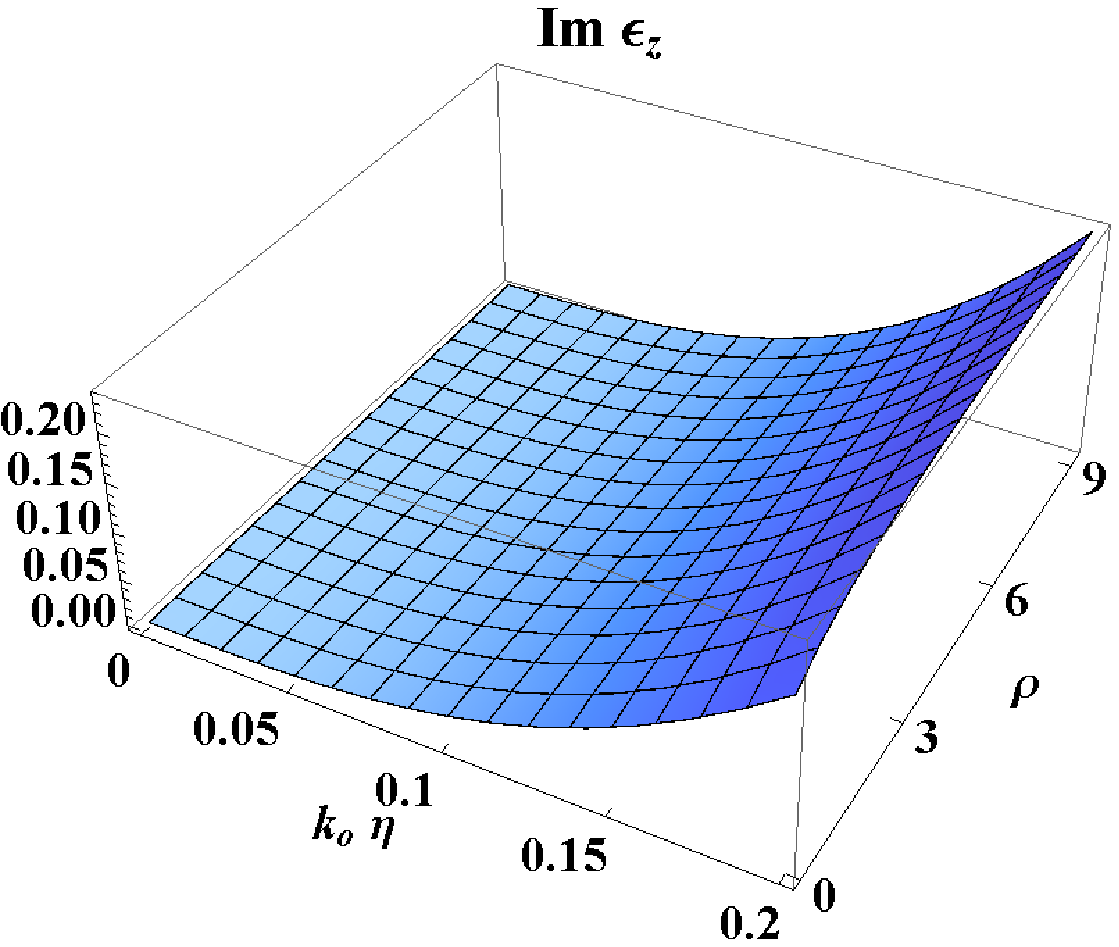,width=2.2in} \\
\epsfig{file=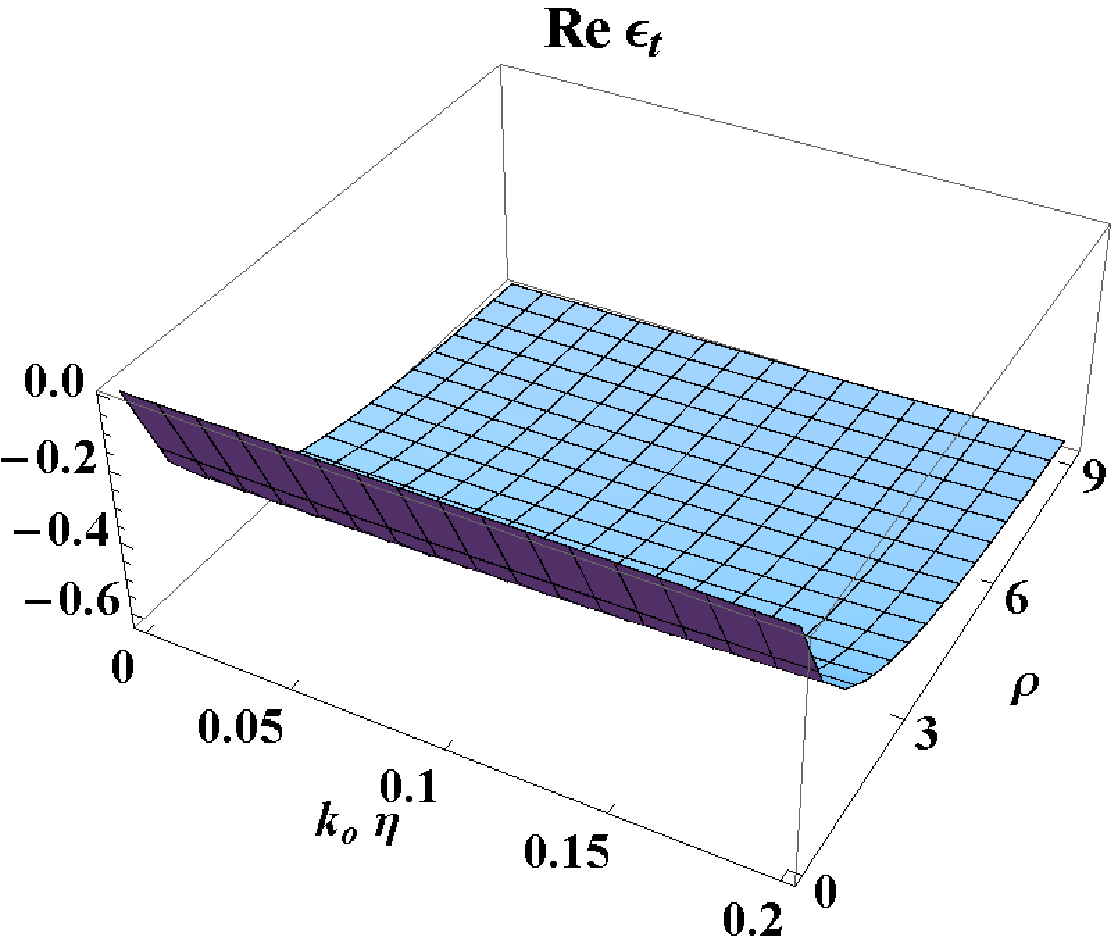,width=2.2in} \hspace{20mm}
\epsfig{file=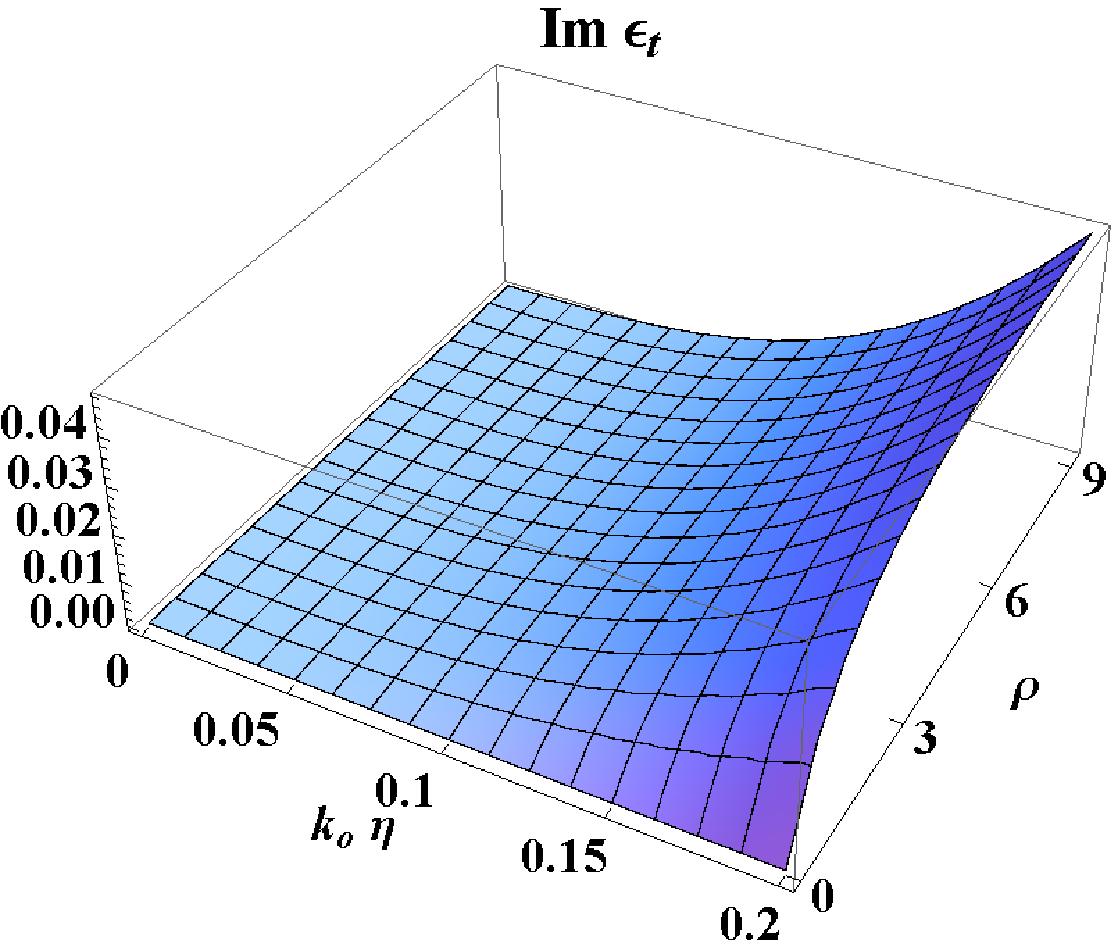,width=2.2in}
 \caption{As Fig.~\ref{fig1} except that the HCM's relative permittivity
 parameters are plotted versus the relative size parameter $\ko \eta \in \le 0,  0.2 \ri$ and the eccentricity
parameter $\rho \in \le 0, 9 \ri$. The volume fraction $f_a = 0.25 $
 and spheroid orientation angle $\varphi = 45^\circ $.} \label{fig2}
\end{figure}

Before presenting in detail the results of our numerical study of
Voigt wave propagation in the HCM characterized in Figs.~\ref{fig1}
and \ref{fig2}, the following  should be pointed out. In general,
for a given HCM the Voigt wave conditions $Y=0$ and $W \neq 0$ are
satisfied at two distinct orientations, specified by the angular
coordinates $\alpha = \alpha_{1,2}$ and $\beta = \beta_{1,2}$.
However, for the numerical investigations reported  herein we found
that $\alpha_1 \approx \alpha_2$ and $\beta_1 \approx \beta_2$, with
the differences between the two orientations being at most
approximately $ 1\%$ and often much less. Consequently, each curve
of $\alpha$, corresponding to a zero of $|Y|$, presented in the
following Figs.~\ref{fig3}, \ref{fig6} and \ref{fig7}
 would appear
 at higher resolution  as two very closely-spaced curves, namely that of $\alpha_1 $ and that
 of $\alpha_2$.
And similarly for each  curve of $\beta$, and the corresponding
curves of $|W|$, in Figs.~\ref{fig3}, \ref{fig6} and \ref{fig7}.
  However, at  the resolution of Figs.~\ref{fig3}, \ref{fig6} and \ref{fig7}
  this is difficult to
 perceive.  This matter
is elaborated upon in due course,  in the discussion of
Figs.~\ref{fig4} and \ref{fig5}.

Let us now begin by considering the effects  of volume fraction
$f_a$, and spheroid orientation angle $\varphi$, on Voigt wave
propagation in the HCM characterized in Figs.~\ref{fig1} and
\ref{fig2}.
 The angular
coordinates $\alpha$ and $\beta$ of the zeros of $|Y|$, along with
the corresponding values of $|W|$, are plotted versus $f_a \in \le
0, 1 \ri$ in Fig.~\ref{fig3} for $\varphi = 30^\circ$, $60^\circ$
and $90^\circ$. The $\alpha$ coordinate for $\varphi = 90^\circ$
changes abruptly at $f_a = 0.48$, from being very close to
$90^\circ$ for $f_a < 0.48$ to being very close to $180^\circ$ for
$f_a > 0.48$. The change in the $\alpha $ coordinate for mid-range
values of $f_a$ becomes progressively less abrupt as $\varphi$
decreases: for $\varphi = 60^\circ$ the corresponding graph of
$\alpha$ is approximately sigmoidal while for $\varphi = 30^\circ$
the graph is nearly linear. The $\beta$ coordinate is also highly
sensitive to the volume fraction. Indeed, for $\varphi = 90^\circ$,
the value of $\beta $ ranges from $0^\circ$ at $f_a = 0.48$ to
$90^\circ$ in the limits $f_a \to 0$ and $1$. This sensitivity
becomes less pronounced as $\varphi $ decreases. From the
corresponding graphs of $|W|$, we can deduce for which values of
$f_a$ the HCM  supports Voigt wave propagation. In the limits $f_a
\to 0$ and $1$ we see that $|W| \to 0$ and therefore the HCM cannot
support Voigt waves in these limits. There are three further points
for $\varphi = 90^\circ$, namely $f_a = 0.2$, $0.48$ and $0.83$, at
which $|W| =0$ and the HCM cannot support Voigt wave propagation.

As mentioned earlier, each  curve in Fig.~\ref{fig3} (and indeed in
Figs.~\ref{fig6} and \ref{fig7}) actually represents two
closely-spaced but distinct directions for Voigt wave propagation.
In order to better appreciate this feature, in Fig.~\ref{fig4} the
curves of $\alpha_1$ and $\alpha_2$ for $\varphi = 90^\circ$ in
Fig.~\ref{fig3}  are reproduced at much higher resolution. The
corresponding curves of $\beta_{1,2}$ are indistinguishable even at
this higher resolution. At values of $f_a < 0.48$, the graphs of
$\alpha_1$ and $\alpha_2$ are mirror images with respect to
reflection about $\alpha_{1,2} = 90^\circ$, and the same applies to
 the $\alpha_{1,2}$ curves for $f_a > 0.48$ but with respect to the
symmetry line $\alpha_{1,2} = 180^\circ$. The differences between
the $\alpha_1$ and $\alpha_2$ curves ranges from approximately
$1^\circ$ down to $0^\circ$. Notice that the $f_a$  values at which
the difference is $0^\circ$, i.e., $f_a = 0.2$ and $0.83$,
correspond to the zeros of $|W|$ observed in Fig.~\ref{fig3}.

In order to illustrate the two distinct orientations for Voigt wave
propagation for $\varphi \neq 90^\circ$, in Fig.~\ref{fig5} a
representative example is provided for $\varphi = 30^\circ$, with
$f_a = 0.3$ and the eccentricity and size parameters being the same
as in Fig.~\ref{fig3}. Here the normalized values of $|Y|$ are
plotted versus the angular coordinates $\alpha \in \le 164.5^\circ,
165.5^\circ \ri$ and $\beta \in \le 70.4^\circ, 71.4^\circ \ri$. Two
$|Y|$ zeros can be seen: one at $\alpha = 164.8^\circ$, $\beta =
70.9^\circ$ and the other at $\alpha = 165.2^\circ$, $\beta =
70.9^\circ$.

Next  the influence of the component particles' shape on the
propagation of Voigt waves is considered.
 The angular
coordinates $\alpha$ and $\beta$ of the zeros of $|Y|$, along with
the corresponding values of $|W|$, are plotted as functions of the
eccentricity parameter $\rho \in \le 0, 9 \ri$ in Fig.~\ref{fig6}
for the spheroid orientation angle $\varphi \in \lec 30^\circ,
60^\circ, 90^\circ \ric$. The volume fraction was fixed at $f_a =
0.25$ and the size parameter $\eta = 0.2 /\ko$.
 The values of  $\alpha$ for each value of $\varphi$
are quite different, but these values are almost independent of
$\rho$. The values of  $\beta$ for each value of $\varphi$ are also
quite different, but for the values of this angular coordinate
decrease gradually as $\rho$ increases. The  corresponding values of
$|W|$ increase  as $\rho$ increases, most rapidly for $\varphi =
90^\circ$ and least rapidly for $\varphi = 30^\circ$. Furthermore,
$|W| \to 0$ as $\rho \to 0$,  regardless of $\varphi$, as would be
expected since the HCM becomes an isotropic dielectric material in
this limit.

Lastly, we turn to the effect of the component particle size, as
accommodated by the extended Bruggeman formalism. In
Fig.~\ref{fig7}, graphs of the angular coordinates $\alpha$ and
$\beta$ at which $|Y| = 0$, and the corresponding values of $|W|$,
against the relative size parameter  $\ko \eta \in \le 0 , 0.2 \ri$
are displayed for the spheroid orientation angle $\varphi \in \lec
30^\circ, 60^\circ, 90^\circ \ric$. Here the volume fraction $f_a =
0.25$ and the eccentricity parameter $\rho = 9$. While the size
parameter $\eta$ has very little influence upon the orientations for
Voigt waves, as indicated by the nearly horizontal graphs of
$\alpha$ and $\beta$, it does have a profound influence upon whether
or not Voigt waves can propagate. Since $|W| \to 0$ as $\eta \to 0$,
we infer that the nondissipative HCM~---~arising from
vanishingly-small component particles~---~does not support Voigt
wave propagation. In contrast, for nonzero values of $\eta$ we see
that $|W|> 0$ and therefore Voigt wave propagation is supported.

\section{Closing remarks}

Engineered materials, in the form of HCMs, may be conceptualized
which support Voigt wave propagation while their component materials
do not. The case considered here involved remarkably simple
component materials, namely nondissipative isotropic dielectric
materials, in contrast to earlier studies on Voigt-wave-supporting
HCMs which involved dissipative uniaxial dielectric components
\c{ML_Voigt1,ML_Voigt2}. In the present case, the ability of the
HCMs to support Voigt waves relied upon the shape, orientation and
nonzero size of the component particles, as well as the volume
fraction of the component materials. In order to cater for such
component materials, an extended version \c{M_WRM} of the
well-established Bruggeman homogenization formalism
\c{Goncharenko,EAB} was needed. We note that the homogenization
approaches adopted  in earlier Voigt wave studies, to wit the
Maxwell-Garnett formalism \c{ML_Voigt1} and
strong-permittivity-fluctuation theory \c{ML_Voigt2}, could not be
used here because they cannot accommodate  component particles with
differing orientations or component particles of nonzero size.

 Our numerical investigations revealed that the directions in which
 Voigt waves may propagate is highly sensitive to the orientation
 of the component particles and to the volume fraction of the
 component materials, but  less sensitive to the shape of the
 component particles and  less sensitive still to the size of the
 component particles. Furthermore, whether or not a HCM supports
 Voigt wave propagation is critically dependent upon the size of the
 component particles and, in certain cases, upon the volume
 fraction.
For the scenarios considered at here, it was found that, in general,
the two directions which support Voigt wave propagation are very
close together. This is consequence of the component materials being
nondissipative. For dissipative component materials, these two
directions can be  more widely spaced \c{ML_Voigt2}.

This study further emphasizes the role of the micro- and/or
nano-structure in determining the macroscopic optical properties of
engineered materials, and paves the way for a  study of Voigt wave
propagation in bianisotropic HCMs, which~---~courtesy of their
enormous parameter space~---~present greater scope for Voigt waves.


\newpage

\newpage

\newpage

\begin{figure}[!h]
\centering \psfull  \epsfig{file=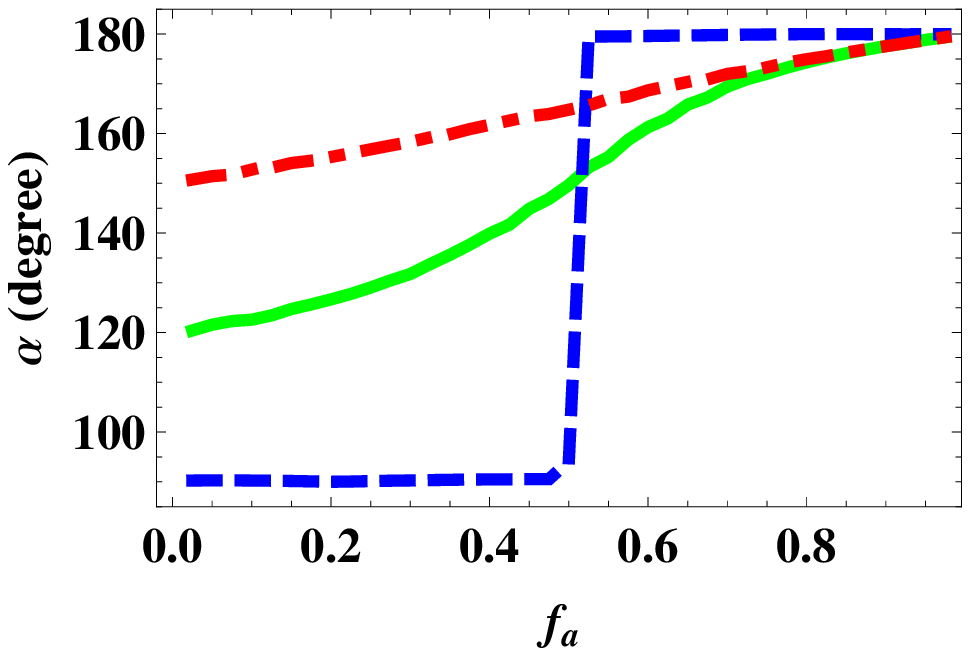,width=3.8in}\\
\epsfig{file=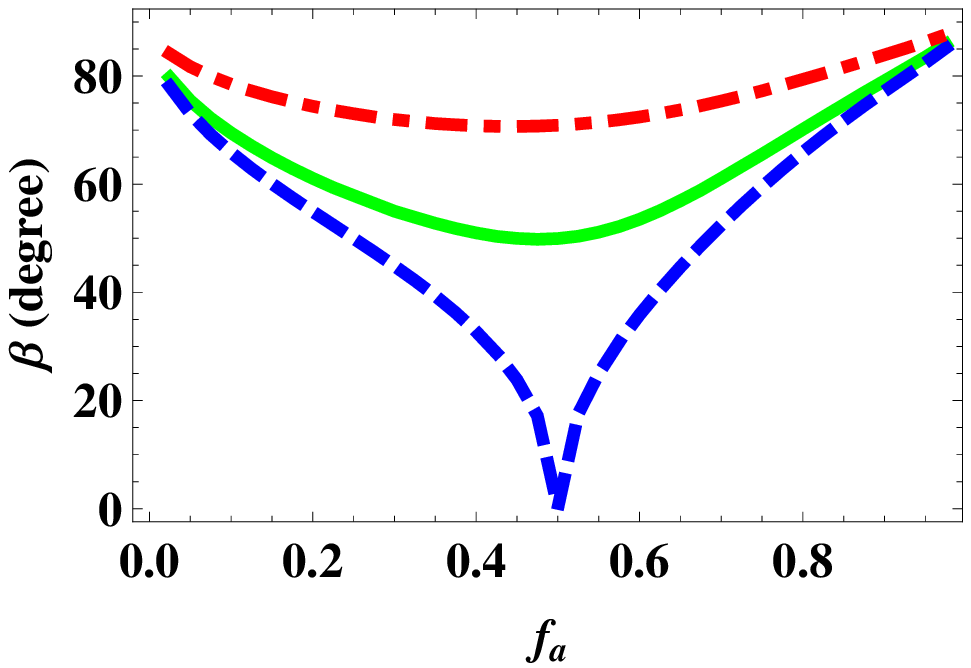,width=3.8in}
\\
\epsfig{file=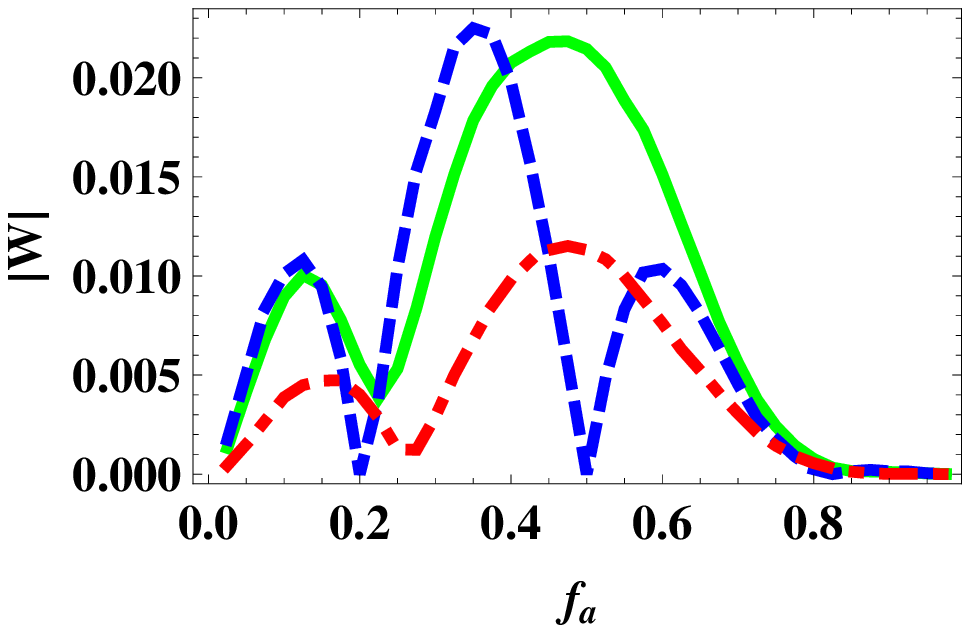,width=3.8in} \caption{The angular
coordinates $\alpha_{}$, $\beta_{}$, and the  absolute value of the
quantity $W_{}$, plotted versus volume fraction $f_a $ for spheroid
orientation angle $\varphi =  90^\circ $ (blue, dashed curves),
$60^\circ$ (green, solid curves) and $30^\circ$ (red, broken dashed
curves). The size parameter $\eta = 0.2 /\ko$ and the eccentricity
parameter $\rho = 9$.
 } \label{fig3}
\end{figure}

\newpage

\begin{figure}[!h]
\centering \psfull
\epsfig{file=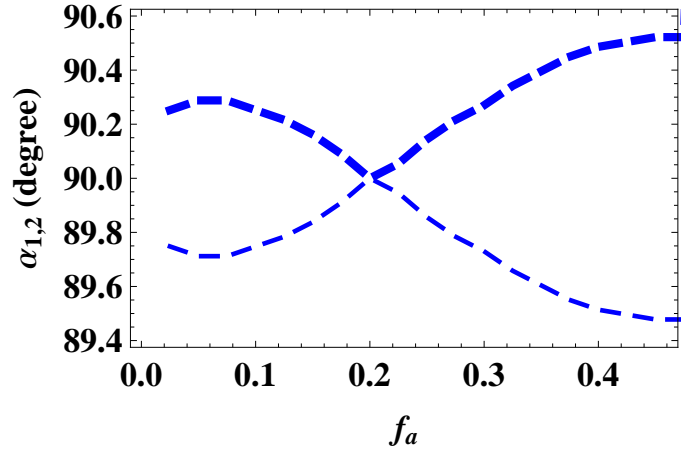,width=3.8in}\\
\epsfig{file=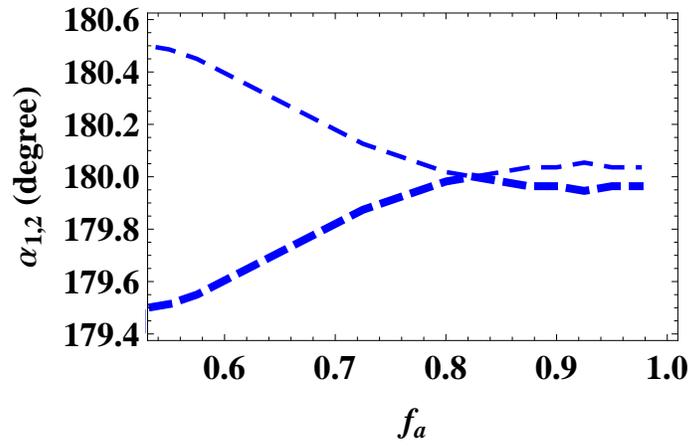,width=3.8in} \caption{Close-up
plots of $\alpha_{}$ versus $f_a$ in Fig.~\ref{fig3} for $\varphi =
90^\circ$. } \label{fig4}
\end{figure}

\newpage

\begin{figure}[!h]
\centering \psfull \epsfig{file=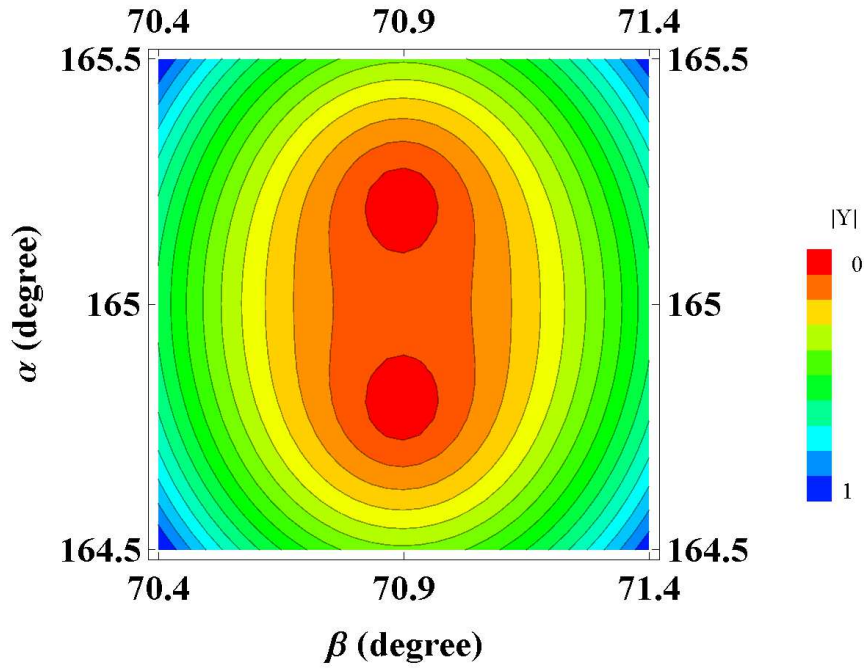,width=4.8in}
\caption{The normalized values of $| Y |$ plotted versus angular
coordinates $\alpha \in \le 164.5^\circ, 165.5^\circ \ri$ and $\beta
\in \le 70.4^\circ, 71.4^\circ \ri$ for spheroid orientation angle
$\varphi = 30^\circ$, volume fraction $f_a = 0.5$, eccentricity
parameter $\rho = 9$ and size parameter $\eta = 0.2 / \ko$. }
\label{fig5}
\end{figure}

\newpage

\begin{figure}[!h]
\centering \psfull  \epsfig{file=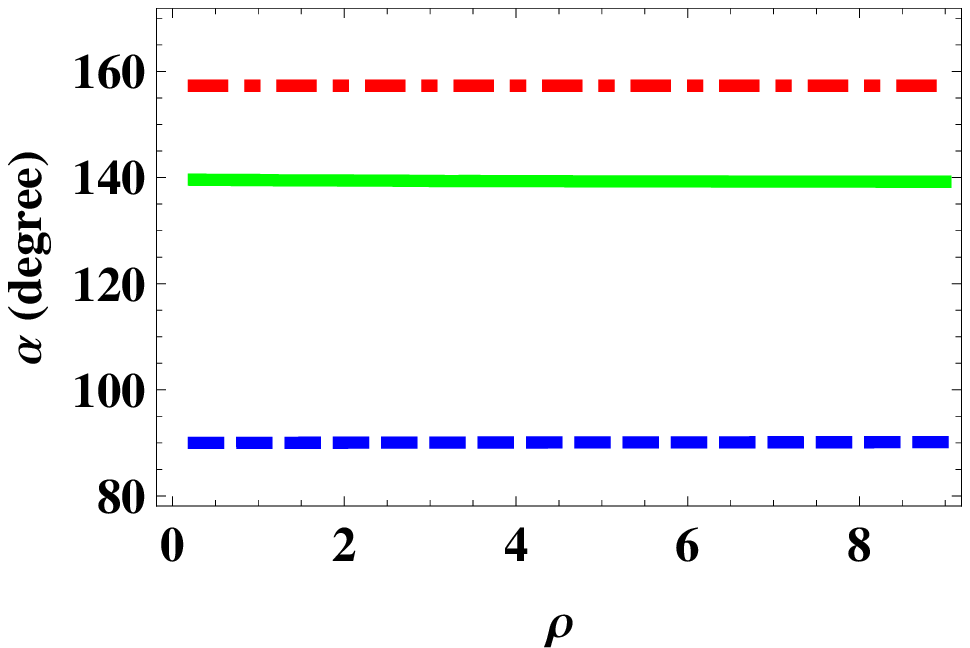,width=3.8in}\\
\epsfig{file=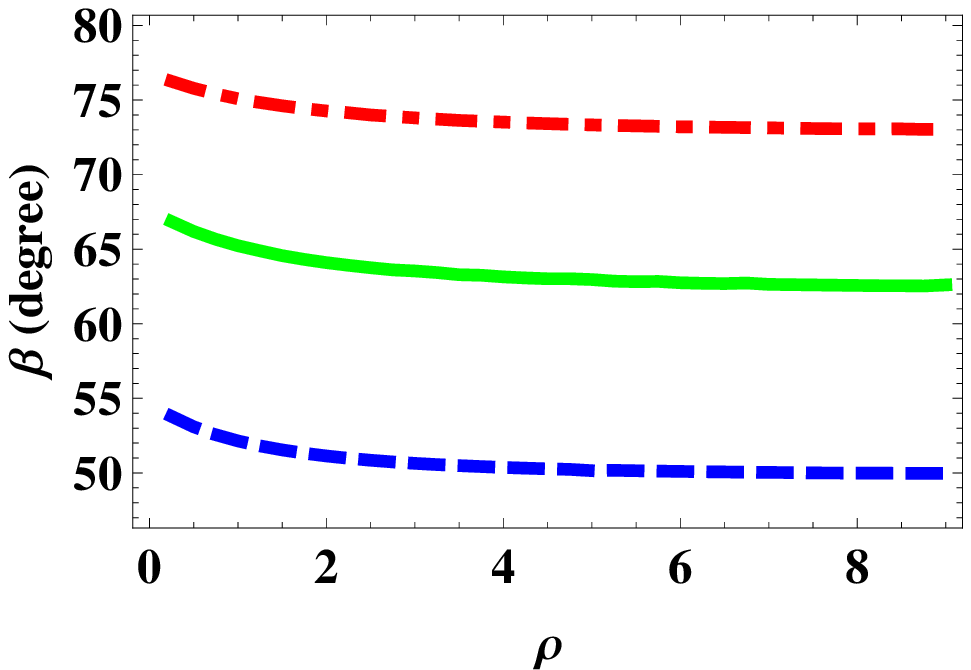,width=3.8in}
\\
\epsfig{file=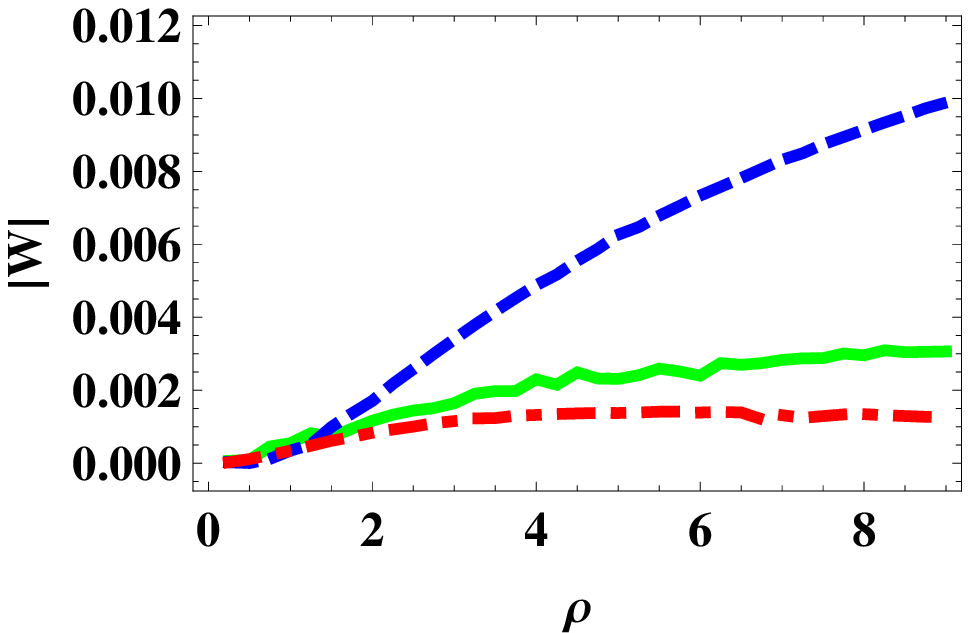,width=3.8in} \caption{The angular
coordinates $\alpha_{}$, $\beta_{}$, and the  absolute value of the
quantity $W_{}$, plotted versus the
 eccentricity parameter $\rho $ for
spheroid orientation angle $\varphi =  90^\circ $ (blue, dashed
curves), $60^\circ$ (green, solid curves) and $30^\circ$ (red,
broken dashed curves). The  size parameter $\eta = 0.2 /\ko$ and the
volume fraction $f_a = 0.25$.
 } \label{fig6}
\end{figure}

\newpage

\begin{figure}[!h]
\centering \psfull  \epsfig{file=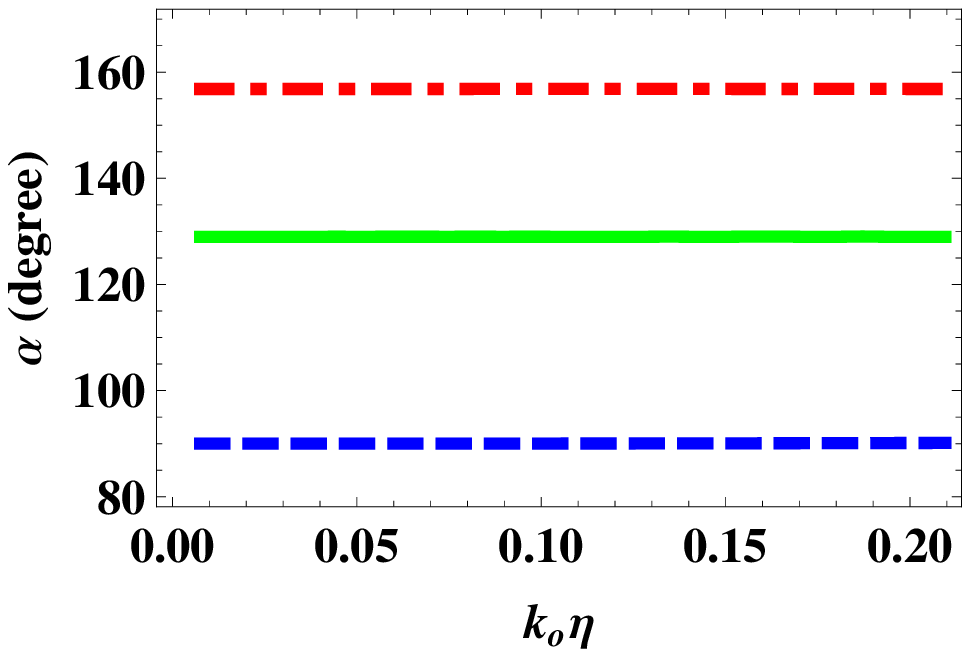,width=3.8in}\\
\epsfig{file=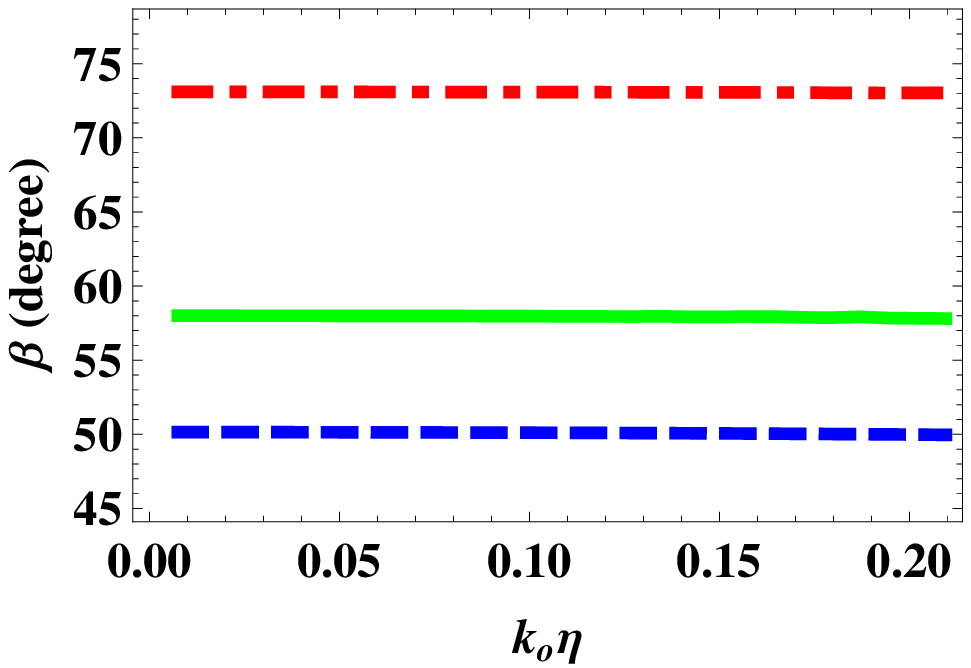,width=3.8in}
\\
\epsfig{file=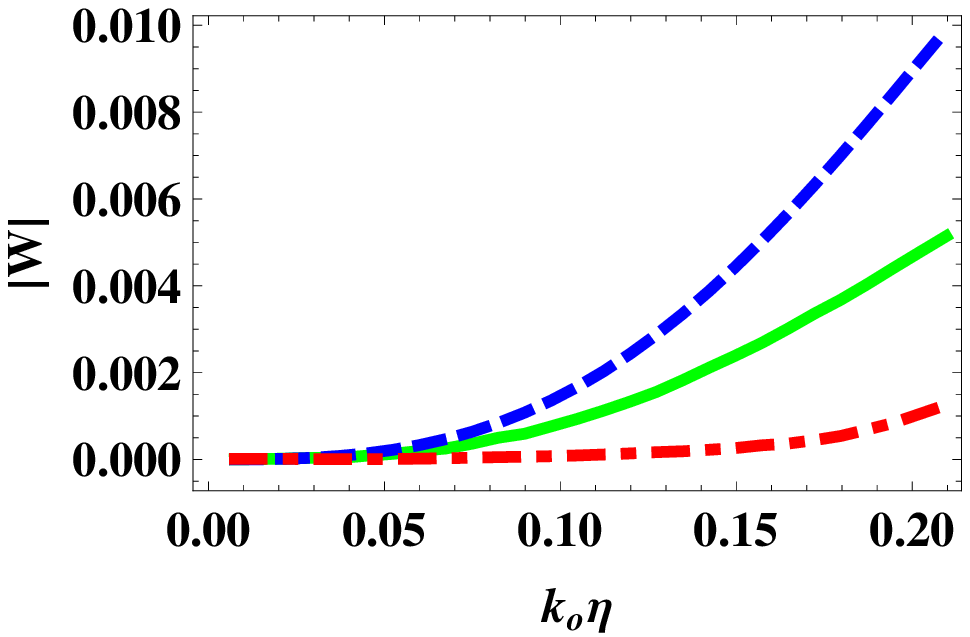,width=3.8in} \caption{The angular coordinates
$\alpha_{}$, $\beta_{}$, and the  absolute value of the quantity
$W_{}$, plotted versus relative size parameter $\ko \eta $ for
spheroid orientation angle $\varphi =  90^\circ $ (blue, dashed
curves), $60^\circ$ (green, solid curves) and $30^\circ$ (red,
broken dashed curves). The volume fraction $f_a = 0.25 $ and the
eccentricity parameter $\rho = 9$.
 } \label{fig7}
\end{figure}

\end{document}